%% file: multispec.tex
\documentclass[a4paper,11pt]{article}
\pdfoutput=1 % to allow pdflatex compilation in JCAP
\usepackage{jcappub}
\usepackage{amsmath}
\usepackage{graphicx}
\usepackage{latexsym}
\usepackage{xspace}
\usepackage{color}
\usepackage{hyperref}
\usepackage{bm}
\usepackage{relsize}
\usepackage{tabularx}
\usepackage{multirow}
\usepackage{amssymb}
\usepackage[table]{xcolor}
\makeatletter
\gdef\@fpheader{}
\g@addto@macro\bfseries{\boldmath}
\makeatother

\input{newcommands}

\subheader{}

\title{Multiple spectator condensates \\ from inflation}

\author{Robert J. Hardwick}

\affiliation{Institute of Cosmology \& Gravitation, University of Portsmouth, Dennis Sciama Building, Burnaby Road, Portsmouth, PO1 3FX, United Kingdom}

\emailAdd{robert.hardwick@port.ac.uk}

\abstract{We investigate the development of spectator (light test) field condensates due to their quantum fluctuations in a de Sitter inflationary background, making use of the stochastic formalism to describe the system. In this context, a condensate refers to the typical field value found after a coarse-graining using the Hubble scale $H$, which can be essential to seed the initial conditions required by various post-inflationary processes. We study models with multiple coupled spectators and for the first time we demonstrate that new forms of stationary solution exist (distinct from the standard exponential form) when the potential is asymmetric. Furthermore, we find a critical value for the inter-field coupling as a function of the number of fields above which the formation of stationary condensates collapses to $H$. Considering some simple two-field example potentials, we are also able to derive a lower limit on the coupling, below which the fluctuations are effectively decoupled, and the standard stationary variance formulae for each field separately can be trusted. These results are all numerically verified by a new publicly available python class (\href{https://github.com/umbralcalc/nfield}{\texttt{nfield}}) to solve the coupled Langevin equations over a large number of fields, realisations and timescales. Further applications of this new tool are also discussed.}


\begin{document}
\sloppy

%\arxivnumber{16XX.XXXXX}

\maketitle

\section{Introduction}
\label{sec:intro}
An effectively massless (light), energetically sub-dominant (test) scalar field placed in an inflationary background is one of the simplest models in which one investigates Quantum Field Theory (QFT) in curved spacetime. Reflecting this simplicity, light test fields --- often termed `spectator' fields --- have been studied with great detail in the literature~\cite{Starobinsky:1994bd,Linde:1996gt,Enqvist:2001zp,Lyth:2001nq, Moroi:2001ct,Bartolo:2002vf,Enqvist:2012xn,Hardwick:2017fjo, Hardwick:2017qcw}.

As a consequence of these quantum effects, spectator fields typically form condensates during inflation, where here a `condensate' refers to the root mean squared of the distribution over field values in each Hubble patch which one can take to be the typical field value. Such condensates can be used to set the initial conditions for subsequent epochs, e.g., reheating, which connects the inflationary sector to the Standard Model (SM). These initial conditions can therefore depend on the potentials of both the spectator and the inflaton, and hence it is quite natural to consider valid extensions to the standard cases considered in the literature for distinct post-inflationary predictions.

In the same vein, consider now the evolution of multiple spectator fields in the inflationary background. Fields of this type are known to exhibit logarithmic divergences in their correlation functions during inflation, which arise from the accumulation of super-horizon modes sourced by quantum fluctuations that cross the horizon. Due to the unique quantum state~\cite{PhysRevLett.59.2555,
PhysRevD.46.1440,PhysRevD.42.3413} of such modes during any slow-roll phase, a coarse-graining can be performed over the field at a given physical scale\footnote{The coarse-graining scale is typically a Fourier mode $k_{\rm cg}=\epsilon aH$, where $\epsilon$ is a parameter which in principle can be linked to the renormalisation scale of the field~\cite{inprep}. In the small $\epsilon$ limit, the results from stochastic inflation become independent of $\epsilon$~\cite{Grain:2017dqa}.}. This coarse-graining reveals an Infra-Red (IR) behaviour where the non-commutative components of the field become relatively small when compared with their anti-commuting components. The quantum correction to the classical field dynamics can thus be well-described as a stochastic system of drift and diffusion~\cite{Tsamis:2005hd} captured by the following Langevin equation for an indexed field $\sigma_i$ appearing in a multi-field potential $V$
\begin{equation} \label{eq:langevin}
\frac{\dd \sigma_i}{\dd N} = -\frac{1}{3H^2}\frac{\partial V}{\partial \sigma_i} + \frac{H}{2\pi}\xi_i (N) \,,
\end{equation}
where $\xi_i$ is a Gaussian white noise term (without cross-correlation) with a unit amplitude ensemble-average $\left\langle \xi_i (N)\xi_j (N') \right\rangle = \delta_{ij}\delta (N-N')$ and the dimensionless time variable is the number of \efold{s} $N \equiv \ln a$ ($a$ being the scale factor). In all equations throughout this paper, we will use the indices $i,j,k =\{ 1,2,\dots ,n_{\rm f}\}$, where $n_{\rm f}$ is the number of spectator fields.

We note here that the correlator form of the noise term in \Eq{eq:langevin} originates from the effectively massless and uncoupled mode functions derived from the vacuum solutions to the field in a quasi-de Sitter background. Should the effective mass $\partial^2V/\partial \sigma_i^2$ of the field $\sigma_i$ exceed the Hubble rate, then this formalism is no longer valid and other methods must be developed~\cite{Bunch:1978yq, Birrell:1982ix, Markkanen:2016aes}. Hence, it seems natural here to consider the evolution of light fields up until the threshold where their effective mass is equal to the Hubble rate, and beyond which we shall refer to the condensate as having `collapsed' to the Hubble scale and the effective mass has also saturated to $H$. We shall return to this point in \Sec{sec:comp-analytic-arg} where we, e.g. evaluate the critical couplings required to achieve this saturation.

For interacting fields in de Sitter spacetimes, another critical value is known to exist which signals the breakdown of the semi-classical approximation. In the mean-field approximation, one separates a classical `mean' background field from perturbatively small quantum fluctuations. For quartic scalar fields, in \Ref{Burgess:2010dd}, it was shown that a breakdown in this peturbative expansion occurs in the regime where the bare mass is less than $\lambda H^2/(4\pi^2)$ which cannot be removed by reorganising the perturbative expansion to include a running effective mass. We stress here that non-peturbative methods of resummation, such as those of this paper, are potentially unaffected by such a bound. This is due to the fact that the backreaction from small quantum fluctuations is inherently included into the background evolution described by \Eq{eq:langevin}, thus optimising the perturbative expansion at each new scale in time --- a cosmological analog to (but not exactly the same as~\cite{Woodard:2008yt}) the Renormalisation Group flow~\cite{Tsamis:2005hd}. 

\Eq{eq:langevin} faithfully reproduces the known resummed logarithmic divergences from QFT to leading order~\cite{Tsamis:2005hd, Tokuda:2017fdh}. This is usually referred to as the formalism of Stochastic Inflation~\cite{Starobinsky:1982ee, Starobinsky:1986fx} --- which we shall henceforth refer to as the `stochastic formalism'. The corresponding Fokker-Planck equation to \Eq{eq:langevin} is
\begin{equation} \label{eq:dist}
\frac{\partial }{\partial N}P(\sigma_i ,N) = \frac{1}{3H^2} \sum^{n_{\rm f}}_{j=1}\frac{\partial }{\partial \sigma_j} \left[ \frac{\partial V}{\partial \sigma_j} P(\sigma_i ,N)\right] + \frac{H^2}{8\pi^2}\sum^{n_{\rm f}}_{j=1}\frac{\partial^2}{\partial \sigma_j^2} P(\sigma_i ,N)\,,
\end{equation}
where we have implicitly made use of the test field condition $\partial H /\partial \sigma_i = 0$ and defined $P(\sigma_i ,N)$ as the probability distribution function over field values at a given $N$, when normalised. Thus, the evolution of modes as they accumulate outside of the horizon typically yields an $n_{\rm f}$-dimensional distribution of field displacements throughout the inflationary phase $P(\sigma_i ,N)$.

\Eq{eq:dist} may also be written essentially as a continuity equation~\cite{BIMJ:BIMJ4710280614,Assadullahi:2016gkk}
\begin{equation}
\frac{\partial P}{\partial N} + \sum^{n_{\rm f}}_{i=1}\frac{\partial J_{i}}{\partial \sigma_i} = 0\,,
\end{equation}
where $J_i$ is the probability current and the right hand side of the equation must vanish for probability conservation. By inspection of \Eq{eq:dist}, one may verify that in this case
\begin{equation}~\label{eq:probcurr}
\ J_i = -\frac{1}{3H^2}  \frac{\partial V}{\partial \sigma_i} P(\sigma_i ,N) - \frac{H^2}{8\pi^2}\frac{\partial}{\partial \sigma_i} P(\sigma_i ,N) \,.
\end{equation}

In de Sitter-like inflation the Hubble parameter is effectively constant in time, hence there is a stationary\footnote{$\partial P/\partial N = 0$ in this context.} solution to \Eq{eq:dist}, $P_{\rm stat}$, corresponding to a vanishing divergence $\nabla \cdot \boldsymbol{J} = 0$ of the probability current --- an incompressible flow of the vector field with components $J_i$. Where $n_{\rm f}=1$ in an unbounded field domain\footnote{In the case of a bounded field domain, probability conservation at the specified boundary implies that $J_i=0$ directly.} one can show that in order for the distribution to have a finite normalisation $P(\sigma_1, N) \dd \sigma_1 \rightarrow 0$ (and hence $J_1\rightarrow 0$) as $\sigma_1 \rightarrow \infty$. Furthermore, given $n_{\rm f}=1$, one can also show that in the stationary limit, the vanishing divergence of $J_1$ simply reduces to $\partial J_1 / \partial \sigma_1 = 0$, and $J_1$ must therefore vanish $\forall \sigma_1$. Hence, the left hand side of \Eq{eq:probcurr} may always be set to zero and the well-known exponential solution to \Eq{eq:dist} for the stationary probability distribution is obtained~\cite{Starobinsky:1986fx} $P_{\rm stat} (\sigma_1 ) \propto \exp \left[ -8\pi^2V(\sigma_1)/3H^4\right]$.

For unbounded $V$ with arbitrary $n_{\rm f}$, it is still natural to consider a boundary condition where $J_i=0$ as $\sigma_i\rightarrow \infty$ to restrict unphysical possibilities, and this may even in practice occur at a set finite scale $\Lambda$ that denotes the chosen cutoff of the theory. However, one can no longer generally state that $J_i$ vanishes everywhere throughout the $n_{\rm f}$-field domain since any class of incompressible vector $J_i$ flows are permitted. Because $\partial J_i / \partial \sigma_i = 0$ is still possible, it is true that one stationary solution to \Eq{eq:dist} is
\begin{equation} \label{eq:exp-stat-dist}
\ P_{\rm stat}(\sigma_i ) \propto \exp \left[ -\frac{8\pi^2V(\sigma_i)}{3H^4}\right] \,,
\end{equation}
but it is no longer unique, and one must use either use further analytical arguments or full numerical solutions for verification.

For any $n_{\rm f}$, the stationary distribution $P_{\rm stat}$ is in practice only reached after some equilibration timescale $N_{\rm eq}$. The timescale $N_{\rm eq}$ is defined as the number of \efold{s} it takes for $P(\sigma_i,N)= \prod^{n_{\rm f}}_{i=1}\delta (\sigma_i)$ --- an $n_{\rm f}$-dimensional Dirac delta function\footnote{We note that, for the symmetric potentials about the origin used in this paper, this is of course equivalent to the more general definition of a Dirac function at the global minimum, $\prod^{n_{\rm f}}_{i=1}\delta (\sigma_i - \sigma^{\rm min}_i)$.} --- to relax to $P(\sigma_i,N)=P_{\rm stat}$. Hence, $N_{\rm eq}$ can be thought of as the time it takes for the effective condensate to grow to its maximal value in every field dimension. It is also important to note here that the definition of $N_{\rm eq}$ used in this paper relies on the inflationary background being de Sitter-like. In slow-roll backgrounds where $H$ varies more substantially, such as those permitted by a monomial $U(\phi ) \propto \phi^p$ inflationary potential, this timescale will have to be recomputed~\cite{Hardwick:2017fjo}.

\section{Vanishing probability current with symmetric potentials} \label{sec:proof-symmetric}

In the previous section, we stated that the exponential form (\Eq{eq:exp-stat-dist}) of the stationary solution to \Eq{eq:dist} may no longer be stable when any divergence-free (incompressible) probability currents are potentially allowed. For any choice of $n_{\rm f}>1$, only the divergence of the current must vanish for a stationary solution, which leaves the possibility of a curl in the vector field $\nabla \times \boldsymbol{J}$. Because $\boldsymbol{J}\cdot \hat{\boldsymbol{{\rm e}}}$ vanishes, where $\hat{\boldsymbol{{\rm e}}}$ is the normal to the boundary, the total integral of the curl over the domain of the fields $\sigma_i \in \Sigma$ vanishes according to Stokes' theorem
\begin{equation} \label{eq:stokes}
\int_{\Sigma} (\nabla \times \boldsymbol{J})_i \,\, \dd^{n_{\rm f}}\sigma_i = 0\,,
\end{equation}
however there are still an infinite number of functions for $\nabla \times \boldsymbol{J}$ that can satisfy this criterion. Examining \Eq{eq:probcurr}, and using the general properties of the totally antisymmetric symbol $\epsilon_{ijk}$, one can show that
\begin{equation}
\ ( \nabla \times \boldsymbol{J} )_i  = -\sum^{n_{\rm f}}_{j = 1}\sum^{n_{\rm f}}_{k = 1}\epsilon_{ijk}\frac{1}{3H^2}\frac{\partial V}{\partial \sigma_k}\frac{\partial P}{\partial \sigma_j} \label{eq:curl-J} \,.
\end{equation}
 
Our first remark is that \Eq{eq:curl-J} vanishes at the extrema of $V$ and $P$ (a fact that we numerically verify for a given potential in \Sec{sec:non-vanish}) but not necessarily everywhere in the domain of $\sigma_i$. Secondly, for all choices of potential and initial distribution, if the gradients of $V$ and $P$ align, i.e. $\partial V/\partial \sigma_i \propto \partial P/\partial \sigma_i$, then \Eq{eq:curl-J} vanishes and hence $J_i=0$ must be true at this point. If one takes a derivative of \Eq{eq:exp-stat-dist}, it is clear that the stationary solution that we have quoted satisfies this criterion.

Without an alternative ansatz to compute $P$, it is difficult to make any general claims about stationary solutions to \Eq{eq:dist}, even when $V$ is symmetric\footnote{Indeed, even with symmetric $V$ and $P$ (the latter can be proved to follow from a symmetric initial condition), if $n_{\rm f}=2$ it can be shown that
\begin{equation} \label{eq:arb-func}
\ ( \nabla \times \boldsymbol{J} )_3 = f(\sigma_1,\sigma_2) - f(\sigma_2,\sigma_1) \,,
\end{equation}
where $f(\sigma_1,\sigma_2)$ is an arbitrary function of both variables. \Eq{eq:arb-func} trivially satisfies the integral constraint from Stokes' theorem (\Eq{eq:stokes}) and hence we are left with no further determination of its exact form without working through an explicit example. Note, however, that \Eq{eq:arb-func} gives $( \nabla \times \boldsymbol{J} )_3(\sigma_1,\sigma_2)=-( \nabla \times \boldsymbol{J} )_3(\sigma_2,\sigma_1)$ and hence, if one can also demonstrate that $\boldsymbol{J}(\sigma_1,\sigma_2)=\boldsymbol{J}(\sigma_2,\sigma_1)$ for symmetric potentials, it must be true that $( \nabla \times \boldsymbol{J} )_3=0$. }. However, we conjecture that when $V$ is symmetric, the solution for $P$ --- which we assume has a been evolved from a symmetric initial condition --- typically has a gradient which aligns with $V$ and hence \Eq{eq:exp-stat-dist} is a stable stationary solution to \Eq{eq:dist}. We have verified numerically that \Eq{eq:exp-stat-dist} provides a stable solution to the late-time dynamics with symmetric potentials in \Sec{sec:comp-analytic-arg}. Note also that in asymmetric potentials\footnote{For example, some of the potentials we introduce and discuss in \Sec{sec:dec-coup}.} we can no longer assume that the $J_i$ components vanish everywhere, and \Eq{eq:exp-stat-dist} is no longer the stationary solution. In such instances, one can also turn to numerical methods.

\section{Computation and analytic arguments}
\label{sec:comp-analytic-arg}

The general problem for arbitrary $V$ defined by Eqs. \eqref{eq:langevin} and \eqref{eq:dist} cannot be solved analytically, and so in this section we shall make our computations for the condensates formed from multi-field spectator potentials combining both analytic and numerical methods. The details of our numerical implementation can be found in the Appendix \ref{sec:Num-implement}, where we briefly outline our development of a new publicly available python code, \href{https://github.com/umbralcalc/nfield}{\texttt{nfield}}.

In light of our discussion in \Sec{sec:proof-symmetric}, we cannot always expect to use moments of the distribution in \Eq{eq:exp-stat-dist} to reliably evaluate the stationary variance for asymmetric potentials. However, we shall not need this distribution to hold true in order to still gain an insight from some approximations. 

Consider a general multi-field interacting spectator potential. In the limit of small field displacements, one can typically perform a Taylor expansion about the minimum of a potential which defines an effective mass in each orthogonal field dimension\footnote{Where we have already implicitly performed any necessary rotations in field space such that $\sigma_i\sigma_j$ cross-terms vanish.} as
\begin{equation} \label{eq:effm}
\ M_i^2 \equiv \frac{\partial^2 V}{\partial \sigma_i^2}\,.
\end{equation}
Hence, a generic multi-field potential can be approximated by 
\begin{equation} \label{eq:Veffm}
\ V \simeq \frac{1}{2} \sum^{n_{\rm f}}_{j=1} M_j^2 \sigma_j^2\,,
\end{equation}
where one may account for interactions (both self and with other fields) through the typical values that one finds for $M_i$. For example, quartic self-interacting terms where $M_i\propto \sigma_i$ may be approximately written as $M_i\propto \sqrt{\left\langle \sigma_i^2\right\rangle}$. As another example, consider the situation where $M_i\propto \sigma_k^2$ due to interaction terms, then the effective mass becomes approximately $M_i\propto \left\langle \sigma_k^2\right\rangle$. This approximation will prove sufficient to calculate the desired quantities in \Sec{sec:crit-coup}. 

Using \Eq{eq:Veffm}, one can derive a second-moment evolution equation from \Eq{eq:langevin} of the form~\cite{Starobinsky:1994bd}
\begin{equation} \label{eq:approx-2om}
\frac{\dd \left\langle \sigma^2_i\right\rangle}{\dd N} \simeq -\frac{2M_i^2}{3H^2}\left\langle \sigma_i^2\right\rangle + \frac{H^2}{4\pi^2} \,,
\end{equation}
and hence the stationary\footnote{$\dd \left\langle \sigma_i^2 \right\rangle/\dd N =0$ in this context, hence this need only be `stationary' in the $i$th field dimension. } variance can be immediately derived\footnote{Notice that this equation is indeed consistent with inserting \Eq{eq:Veffm} into \Eq{eq:exp-stat-dist} and taking the second-order moment.}~\cite{Starobinsky:1994bd}
\begin{equation} \label{eq:approx-2om-stat}
\left\langle \sigma_i^2\right\rangle \vert_{\rm stat} = \frac{3H^4}{8\pi^2M_i^2}\,.
\end{equation}
Note that in the limit where \Eq{eq:Veffm} is no longer an approximation, such as for a quadratic non-interacting spectator, then \Eq{eq:approx-2om} and \Eq{eq:approx-2om-stat} are precise equations with $M_i$ corresponding to the bare mass.

\begin{figure}[t]
\begin{center}
\includegraphics[width=0.95\textwidth]{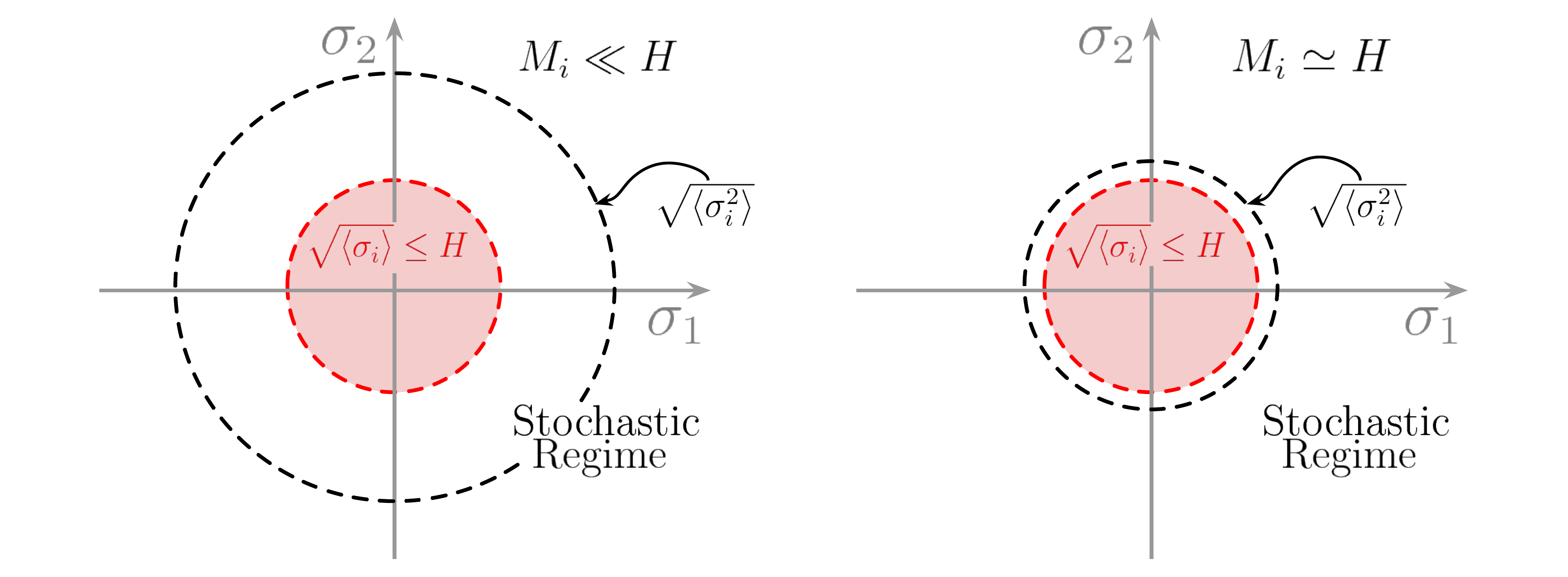}
\caption{~\label{fig:saturation-diagram} Schematic diagrams of the condensate $\sqrt{\left\langle \sigma_i^2 \right\rangle}$ (dashed black circles) in a symmetric two-field potential. When $M_i\simeq H$ the condensate collapses to the Hubble rate, corresponding to the dashed red circle. The shaded red region corresponds to situations where $M_i>H$ and there is conjectured saturation at the Hubble scale $\sqrt{\left \langle \sigma_i^2 \right\rangle } \simeq H$. The entire region in field space outside of the red shaded area can be considered where the stochastic formalism is valid, with the caveat that it is possible for $M_i>H$ to also occur for a field at large displacements, e.g. quartic self-interacting fields will have $M_i \propto \sigma_i$. }
\end{center}
\end{figure}

In \Eq{eq:approx-2om-stat}, the inverse-proportionality between the effective mass and the variance indicates that there is a critical value for $M_i \simeq H$ above which the stationary condensate collapses to the Hubble rate $\sqrt{\left\langle \sigma_i^2\right\rangle} = H$. Taking a two-field example for illustration, we have plotted a schematic diagram of the physical situation in \Fig{fig:saturation-diagram} for a symmetric potential. In the left panel, the condensate (dashed black circle) is relatively large because the effective mass $M_i \ll H$. In the right panel, the condensate collapses to the value of the Hubble rate (dashed red circle) because $M_i\simeq H$. For the shaded red region the suppression $\left\langle \sigma_i^2\right\rangle \propto 1/M_i^2$ in the variance from \Eq{eq:approx-2om-stat} is no longer valid and the stochastic approach can no longer be used. This is because when $M_i>H$ the mode functions which source the fluctuations of the field can no longer be accurately described by the simple form of noise correlator in the definition of \Eq{eq:langevin}.

Other calculations do exist for situations considering a constant super-Hubble mass $M_i>H$~\cite{Bunch:1978yq, Birrell:1982ix, Markkanen:2016aes} where, in these instances, the variance is known to experience further suppression. However, the assumption of constant $M_i$ is one we cannot make for the potentials studied in this work. We anticipate that a similar suppression occurs but leave the verification of this to future work. Even if this is not always true (e.g. for many non-interacting quadratic spectators), the `saturation' value is still of interest since it characterises the fundamental domain of validity for the stochastic formalism. Hence, in this regard, we shall leave the calculation of possible condensates in the $M_i>H$ regime to future work, and therefore we will work within the regime where the stochastic formalism is valid.

\subsection{The critical coupling} \label{sec:crit-coup}

In this subsection, we will demonstrate that when one generalises the formation of spectator field condensates to many coupled fields, a critical value for the coupling appears, above which the equilibrium variances of all fields have collapsed to the Hubble scale and effective mass of each field has saturated to $H$. To show this we will consider a simplified potential that will allow us to calculate this critical coupling both analytically and numerically, for verification.

Now consider the multi-field spectator potential 
\begin{equation} \label{eq:VA}
\ V_{\rm A} = \frac{1}{2} g\sum_{i\neq j} \sigma_i^2\sigma^2_j\,.
\end{equation}
Mindful of the approximation made with $M_i$ in \Eq{eq:Veffm}, one thus expects that incrementally strengthening the interaction between spectator fields $\propto g \sigma_i^2\sigma_j^2$ can lead to the eventual saturation of the condensate value at the Hubble scale due to the effective mass of each field being progressively larger, and we therefore anticipate a critical value for the inter-field coupling $g_{\rm crit}$ to exist, for a given $n_{\rm f}$, above which the stationary condensate collapses to $\sqrt{\left\langle \sigma_i^2\right\rangle} \simeq H$.

By inspection between Eqs. \eqref{eq:Veffm} and \eqref{eq:VA}, the typical value of the effective mass in the $i$th field dimension corresponds to $M_i^2 \simeq g\sum_{k\neq i}\left\langle  \sigma_k^2\right\rangle \simeq g (n_{\rm f}-1)\left\langle  \sigma_i^2\right\rangle $, where in the second equality we have assumed that the distribution (using \Eq{eq:VA} as the potential) has reached stationarity $P(\sigma_i,N) = P_{\rm stat}$ and, hence, due to symmetry $\left\langle \sigma^2_i \right\rangle = \left\langle \sigma^2_k \right\rangle \,\, \forall \,\, i,k$. Because $\left\langle \sigma^2_i \right\rangle = \left\langle \sigma^2_i \right\rangle \vert_{\rm stat}$, given in \Eq{eq:approx-2om-stat}, we can now obtain an approximate relation for the critical coupling\footnote{Note that because \Eq{eq:VA} is a symmetric potential --- i.e. $V(\sigma_{j}) = V(\sigma_{{\rm perm}(j)})$ for any permutation of field indices ${\rm perm}(j)$ --- our discussion in \Sec{sec:proof-symmetric} indicates the stability of \Eq{eq:exp-stat-dist} in this situation. Hence, another way to compute \Eq{eq:gcrit-variance-analytic} would be to take the second moment of \Eq{eq:exp-stat-dist}.}
\begin{align}
\left\langle  \sigma_i^2\right\rangle &= \frac{3H^4}{8\pi^2 M_i^2} \simeq \frac{3H^4}{8\pi^2 g (n_{\rm f}-1)\left\langle  \sigma_i^2\right\rangle } \nonumber \\
&\Rightarrow \left\langle  \sigma_i^2\right\rangle \simeq \sqrt{ \frac{3H^4}{8\pi^2 g (n_{\rm f}-1)}} \label{eq:gcrit-variance-analytic} \\
&\Rightarrow g_{\rm crit} \simeq \frac{3}{8\pi^2 (n_{\rm f}-1)}  \label{eq:gcrit-analytic}\,,
\end{align}
where we have found $g_{\rm crit}$ by setting $\left\langle \sigma_i^2\right\rangle = H^2$ in \Eq{eq:gcrit-variance-analytic}.

\begin{figure}[t]
\begin{center}
\includegraphics[width=0.48\textwidth]{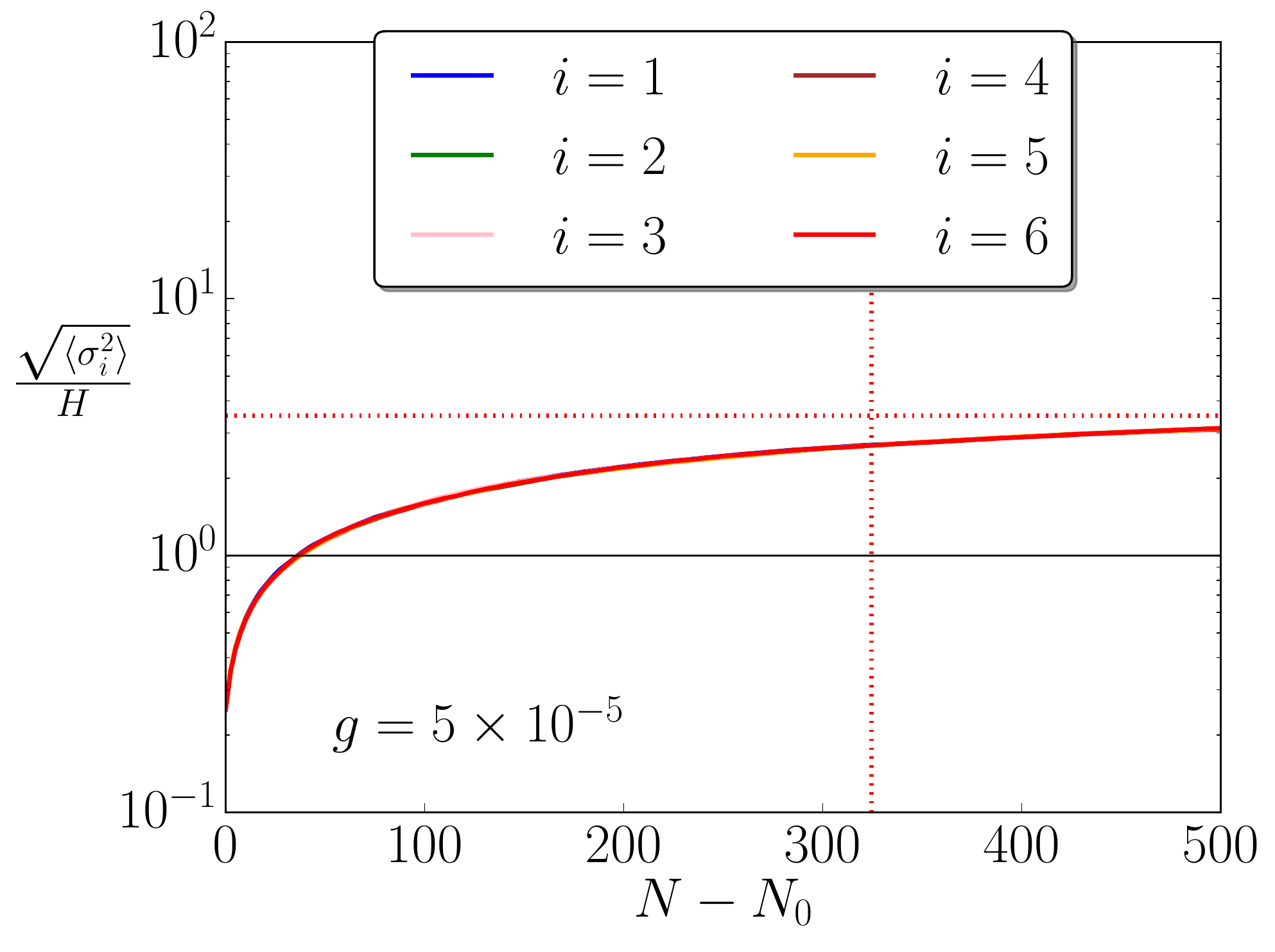}
\includegraphics[width=0.48\textwidth]{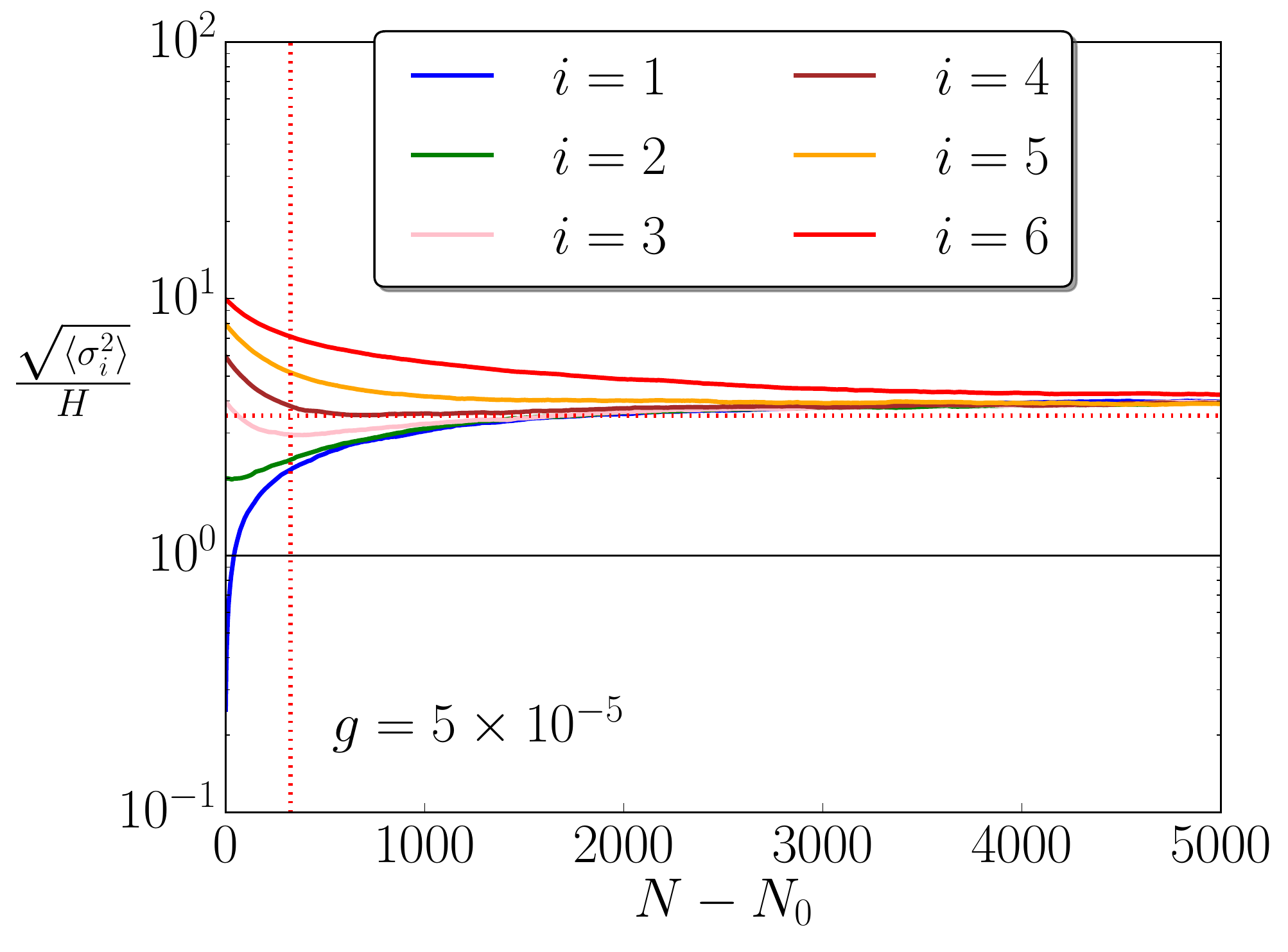}
\caption{~\label{fig:VA-variance-plot} The numerically evaluated (solid lines) time evolution of the variance for each spectator field in the case of the $V_{\rm A}$ potential (\Eq{eq:VA}) with example value $n_{\rm f}=6$. The variance of all fields is initialised at $\left\langle \sigma_i^2\right\rangle = 0$ in the left panel and we have chosen a range of initial conditions for the fields in the right panel to indicate the robustness of the late time stationary behaviour. All of the field variances overlap due to the symmetry of the potential. Dotted horizontal and vertical lines represent the stationary variance (\Eq{eq:gcrit-variance-analytic}) and equilibration timescale (\Eq{eq:Neqi}), respectively. The number of realisations used is $10^4$. }
\end{center}
\end{figure}

\begin{figure}[t]
\begin{center}
\includegraphics[width=0.65\textwidth]{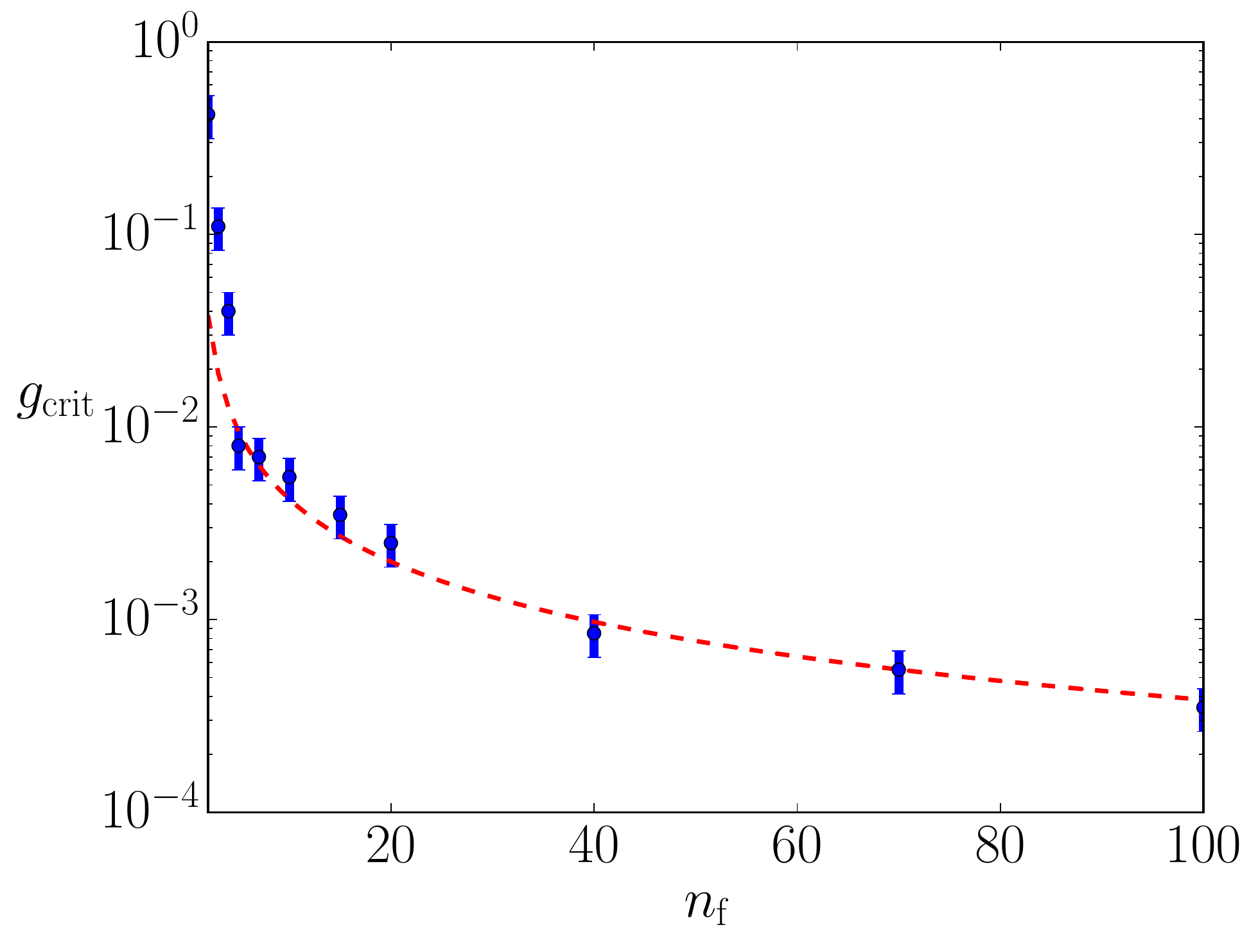}
\caption{~\label{fig:gcrit-plot} The numerically evaluated (blue data points with error bars related to both statistical and numerical uncertainty from having a finite number of realisations and a finite stepsize in numerically finding $g_{\rm crit}$, respectively) value of $g_{\rm crit}$ as a function of the number of fields $n_{\rm f}$ for potential $V_{\rm A}$ (see \Eq{eq:VA}). The line clearly matches the analytically derived relation in \Eq{eq:gcrit-analytic} (dashed red line) very well. The number of realisations used is $10^4$ for each point. Due to how rapidly $g_{\rm crit}$ varies with the number of fields for $n_{\rm f}\lesssim 5$, we are likely underestimating our error in this region, and hence these points may appear slightly inconsistent. }
\end{center}
\end{figure}

For illustration, we plot the time evolution for variances, averaging over multiple Langevin realisations (realisations of \Eq{eq:langevin}), of an example where $n_{\rm f}=6$ in \Fig{fig:VA-variance-plot} and \Eq{eq:gcrit-variance-analytic} is shown to be a good description of the stationary values against the numerically evaluated variances (all identical to each other due to the symmetry). The approximate form of \Eq{eq:gcrit-analytic} must also be verified numerically, and hence we plot in \Fig{fig:gcrit-plot} the comparison between numerical and analytic approaches to obtain the functional relationship between $g_{\rm crit}(n_{\rm f})$. Due to the apparently excellent agreement between the two calculations in \Fig{fig:gcrit-plot} we can be confident in \Eq{eq:gcrit-analytic} as a reliable formula to extrapolate to large $n_{\rm f}$.

We further note that one may derive the equilibration timescale for each field dimension $N_{{\rm eq},i}$ for the $V_{\rm A}$ potential, and this is approximately be given by
\begin{equation} \label{eq:Neqi}
\ N_{{\rm eq},i} \simeq \frac{H^2}{M_i^2} \simeq \sqrt{\frac{8\pi^2}{3g(n_{\rm f}-1)}} \,.
\end{equation}
The first relation of \Eq{eq:Neqi} can be derived from \Eq{eq:approx-2om} (see also Refs.~\cite{Starobinsky:1986fx,Enqvist:2012xn,Hardwick:2017fjo}), where it is also natural to consider the `steepness' of the effective potential in \Eq{eq:Veffm} to control the rate of equilibration. Note that \Eq{eq:Neqi} has been derived by assuming the stationary variance, however, because $N_{\rm eq,i}$ is precisely the time it takes to relax to the stationary limit, this assumption is not strictly valid and requires comparison with full numerical solutions. Interestingly, in the example with $n_{\rm f}=6$ plotted in \Fig{fig:VA-variance-plot}, \Eq{eq:Neqi} appears to perform well regardless of its less trustworthy origin.

\subsection{The decoupling limit} \label{sec:dec-coup}

We will now investigate another limit of the inter-field coupling, which can also be analytically estimated for some specific potentials and the numerical verification will also serve to showcase further applications of the \href{https://github.com/umbralcalc/nfield}{\texttt{nfield}} code.

Consider two further examples of interacting spectator potentials
\begin{align}
\ V_{\rm B} &= \frac{1}{2}m^2\left( \sigma^2_1 + \alpha \sigma^2_2\right) + \frac{1}{2}g \sigma_1^2\sigma^2_2 \label{eq:VBpot} \\
\ V_{\rm C} &= \frac{1}{4}\lambda \left( \sigma^4_1 + \alpha \sigma^4_2\right) + \frac{1}{2}g \sigma_1^2\sigma^2_2\,. \label{eq:VCpot}
\end{align}
$V_{\rm B}$ a generalisation from $V_{\rm A}$ by introducing additional masses $m$ and $\sqrt{\alpha} m$, with a hierarchy parameter $\alpha$, but we have now specified that $n_{\rm f}=2$ to capture the essential phenomenology. $V_{\rm C}$ is another generalisation from $V_{\rm A}$ to include self-interactions. We note here that, in each case, decoupling the system in the limit where $g \rightarrow 0$ will yield the well-known formulae~\cite{Starobinsky:1986fx,Enqvist:2012xn,Hardwick:2017fjo} for the stationary variance of each non-interacting field (see \Eq{eq:approx-2om-stat})
\begin{align}
\ {\rm Decoupled} \,\, V_{\rm B} \Rightarrow \left\langle \sigma_1^2 \right\rangle \vert_{\rm stat} &= \frac{3H^4}{8\pi^2 m^2} \label{eq:stationary_VB_variances_1} \\
\left\langle \sigma_2^2 \right\rangle \vert_{\rm stat} &= \frac{3H^4}{8\pi^2 \alpha m^2} \label{eq:stationary_VB_variances_2} \\
\ {\rm Decoupled} \,\, V_{\rm C} \Rightarrow \left\langle \sigma_1^2 \right\rangle \vert_{\rm stat} &= \frac{\Gamma \left( \frac{3}{4}\right)}{\Gamma \left( \frac{1}{4}\right)}\sqrt{\frac{3H^4}{2\pi^2 \lambda}} \label{eq:stationary_VC_variances_1} \\
\left\langle \sigma_2^2 \right\rangle \vert_{\rm stat} &= \frac{\Gamma \left( \frac{3}{4}\right)}{\Gamma \left( \frac{1}{4}\right)}\sqrt{\frac{3H^4}{2\pi^2 \alpha \lambda}} \label{eq:stationary_VC_variances_2} \,.
\end{align}
In this same limit one can also obtain the respective equilibration timescales
\begin{align}
\ {\rm Decoupled} \,\, V_{\rm B} \Rightarrow N_{{\rm eq},1} \vert_{\rm stat} &\simeq \frac{H^2}{M_1^2} \simeq \frac{H^2}{m^2} \label{eq:stationary_VB_Neq_1} \\
\ N_{{\rm eq},2} \vert_{\rm stat} &\simeq \frac{H^2}{M_2^2} \simeq \frac{H^2}{\alpha m^2} \label{eq:stationary_VB_Neq_2} \\
\ {\rm Decoupled} \,\, V_{\rm C} \Rightarrow N_{{\rm eq},1} \vert_{\rm stat} &\simeq \frac{H^2}{M_1^2} \propto \frac{1}{\sqrt{\lambda}} \label{eq:stationary_VC_Neq_1} \\
\ N_{{\rm eq},2} \vert_{\rm stat} &\simeq \frac{H^2}{M_2^2} \propto \frac{1}{\sqrt{ \alpha \lambda}} \label{eq:stationary_VC_Neq_2} \,,
\end{align}
where we recall that the effective masses $M_i$ are defined in \Eq{eq:effm}. We note there that, as in \Eq{eq:Neqi}, these timescales have been derived using the stationary form of the variances which is not strictly valid, hence they must be checked for validity against the numerical implementation to ensure that they are still accurate.

A `decoupling' value of $g=g_{\rm dec}$ can be derived analytically from these stationary variances by obtaining the value of $g$ above which the main contribution to the effective mass is from the coupling term $\propto g$ and not from the bare mass or self-interaction. Looking at the effective mass of either of the fields in each potential, one can hence show that in the stationary limit
\begin{align}
\ V_{\rm B} &\Rightarrow M^2_1 \simeq m^2 + g \left\langle \sigma_2^2 \right\rangle\vert_{\rm stat} \Rightarrow g_{\rm dec} \simeq \frac{8\pi^2}{3}\alpha \left( \frac{m}{H}\right)^4  \label{eq:gdecVB} \\
\ V_{\rm C} &\Rightarrow M^2_1 \simeq 3\lambda \left\langle \sigma_1^2 \right\rangle \vert_{\rm stat} + g \left\langle \sigma_2^2 \right\rangle\vert_{\rm stat} \Rightarrow g_{\rm dec} \simeq 3 \lambda \sqrt{\alpha} \,. \label{eq:gdecVC}
\end{align}
If one were to re-derive \Eq{eq:gdecVB} and \Eq{eq:gdecVC} by replacing $M_1 \Longleftrightarrow M_2$, it is trivial to show that the same formulae are obtained.

\begin{figure}[t]
\begin{center}
\includegraphics[width=0.45\textwidth]{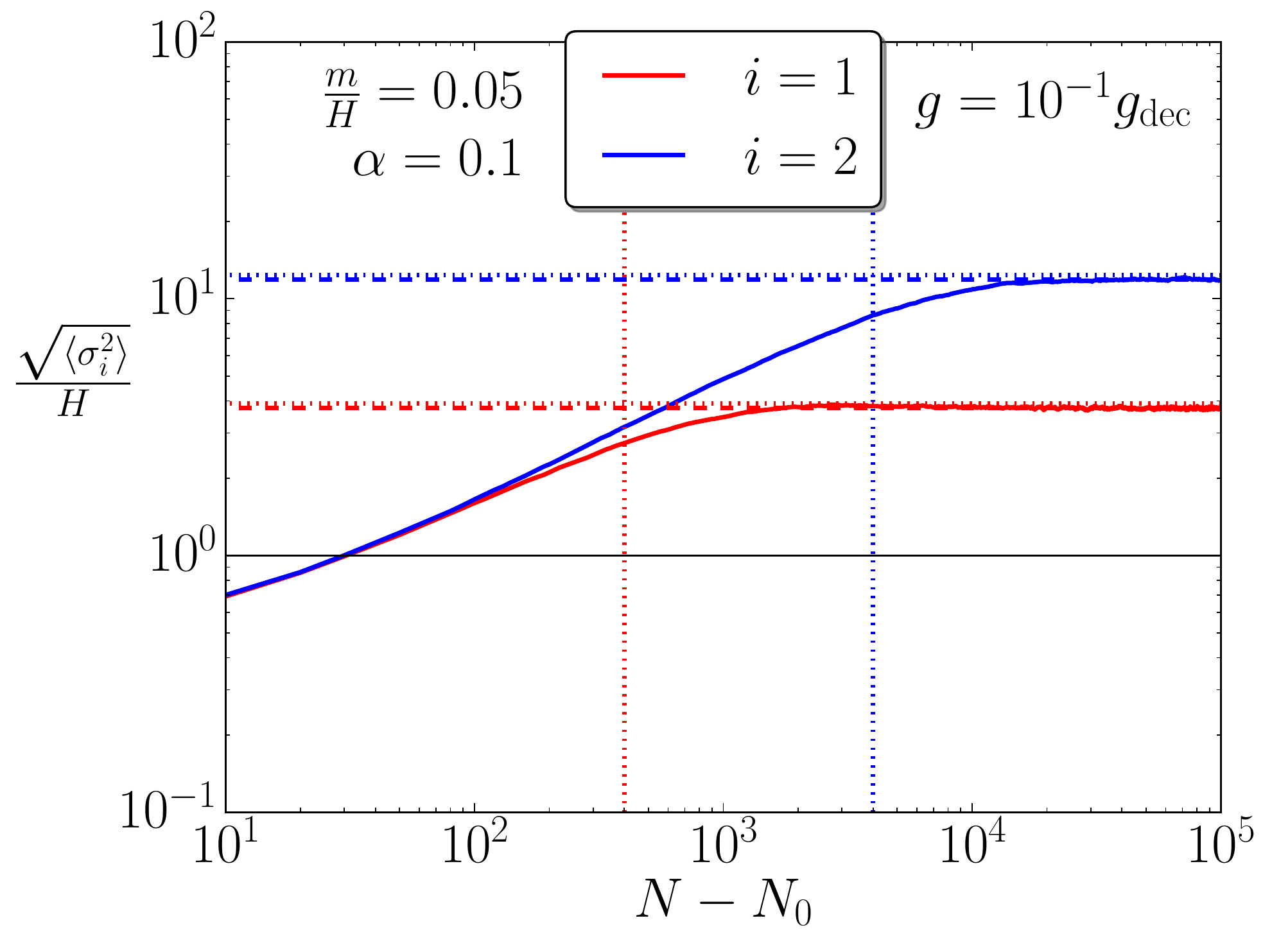}
\includegraphics[width=0.45\textwidth]{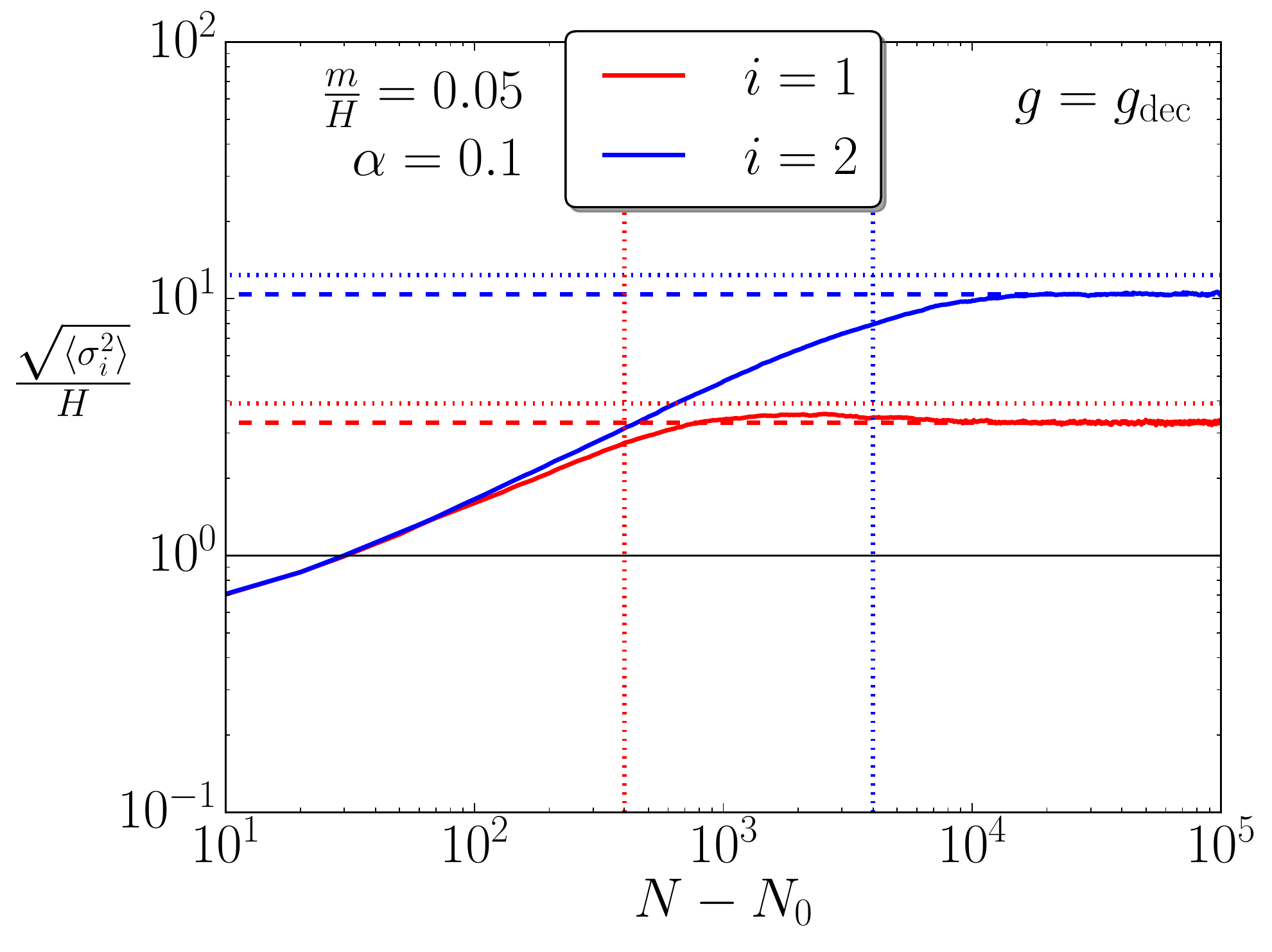} \\
\includegraphics[width=0.45\textwidth]{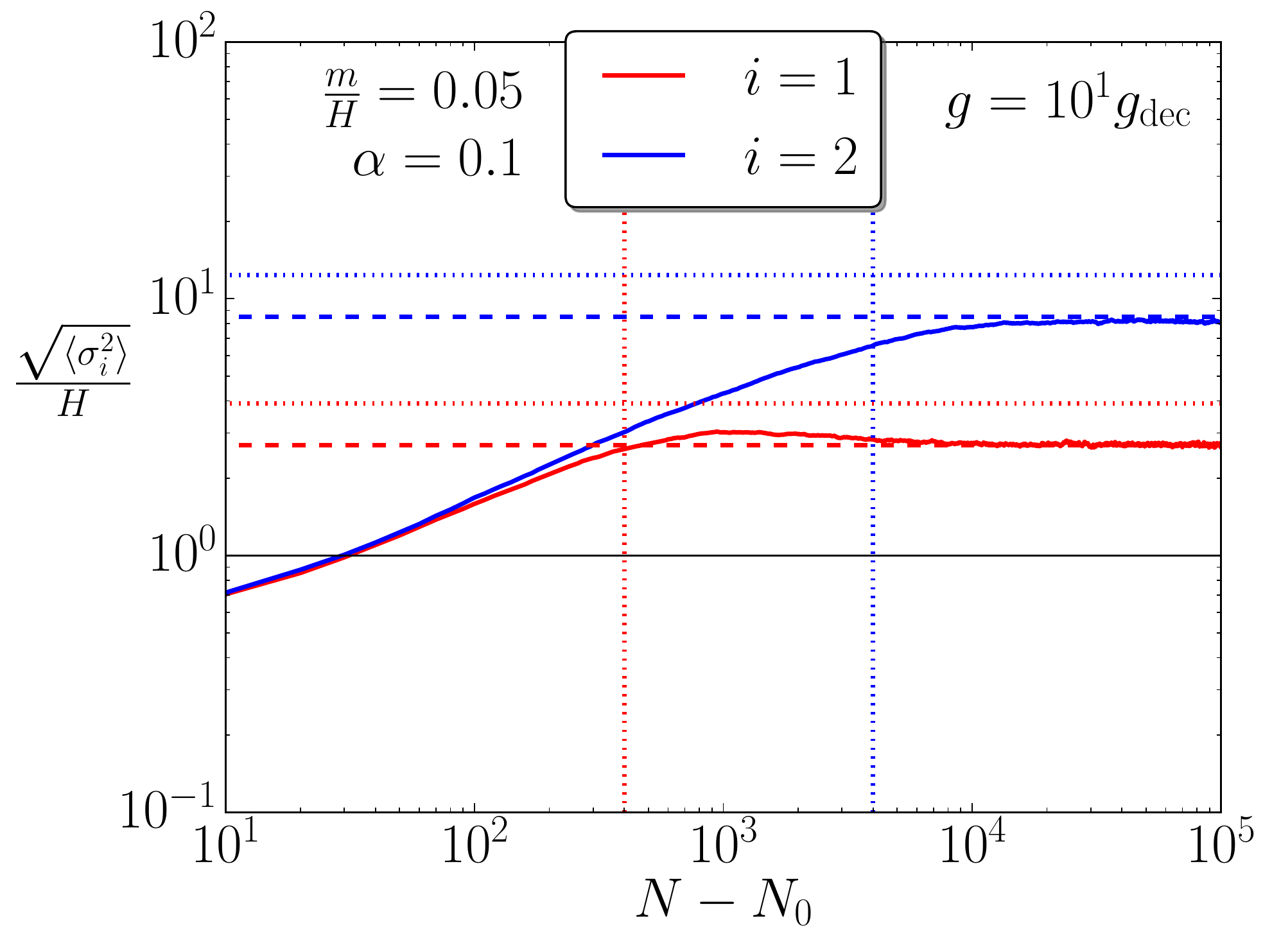}
\includegraphics[width=0.45\textwidth]{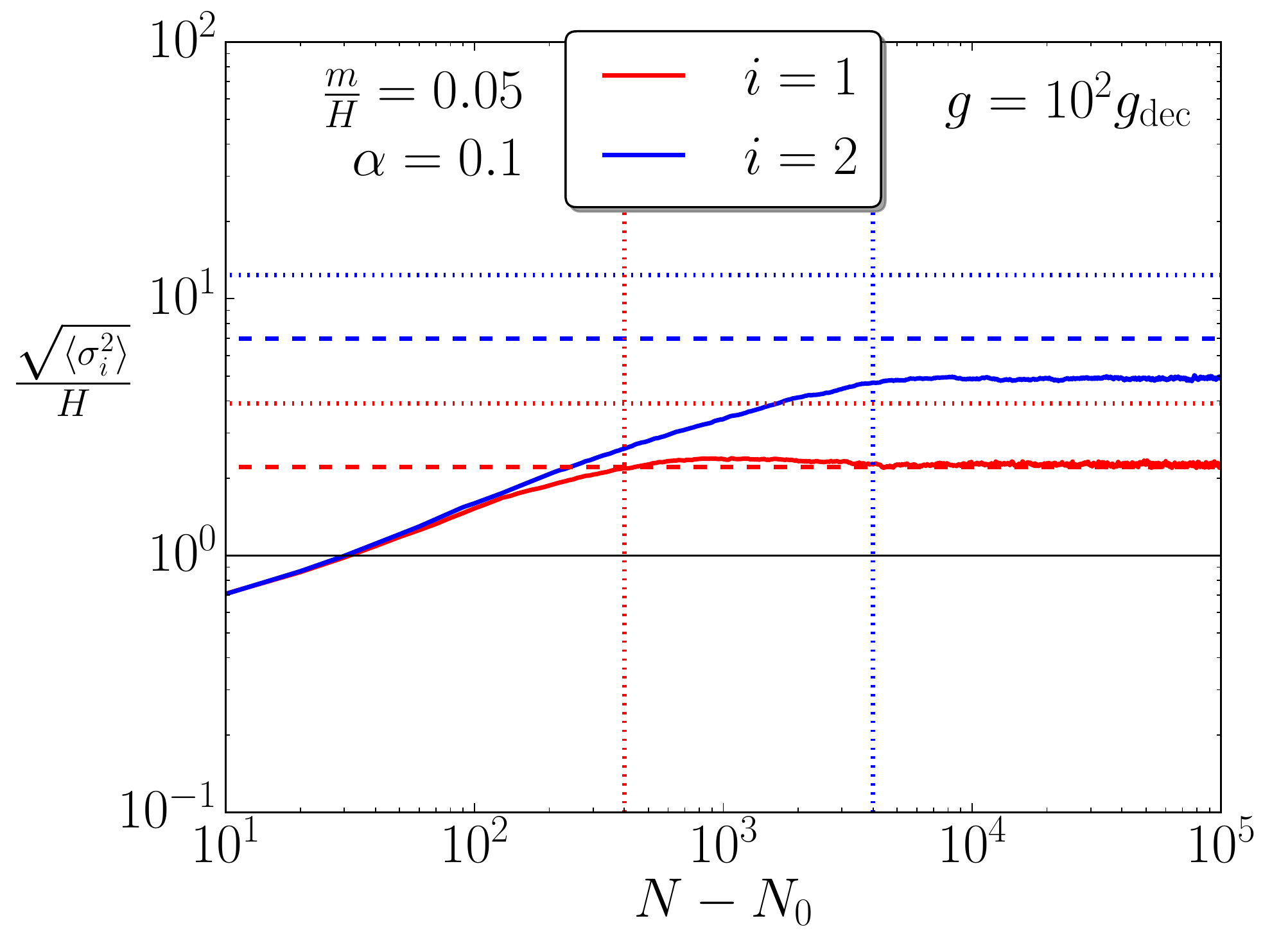}
\caption{~\label{fig:quadratic-plots} The numerically evaluated (solid lines) time evolution of the variance for each spectator field in the case of the $V_{\rm B}$ potential (see \Eq{eq:VBpot}), where variance of both fields is initialised at $\left\langle \sigma_i^2\right\rangle = 0$ and $10^4$ realisations of \Eq{eq:langevin} were used in \href{https://github.com/umbralcalc/nfield}{\texttt{nfield}}. Dotted horizontal and dotted vertical lines represent the stationary variances (Eqs. \eqref{eq:stationary_VB_variances_1} and \eqref{eq:stationary_VB_variances_2}) and equilibration timescales (Eqs. \eqref{eq:stationary_VB_Neq_1} and \eqref{eq:stationary_VB_Neq_2}), respectively, computed in the decoupled limit $g\leq g_{\rm dec}$. The dashed horizontal lines use an alternative method to derive the stationary variance by numerically evaluating the second moment of \Eq{eq:exp-stat-dist}. The number of realisations used in each case is $10^4$. }
\end{center}
\end{figure}

\begin{figure}[t]
\begin{center}
\includegraphics[width=0.45\textwidth]{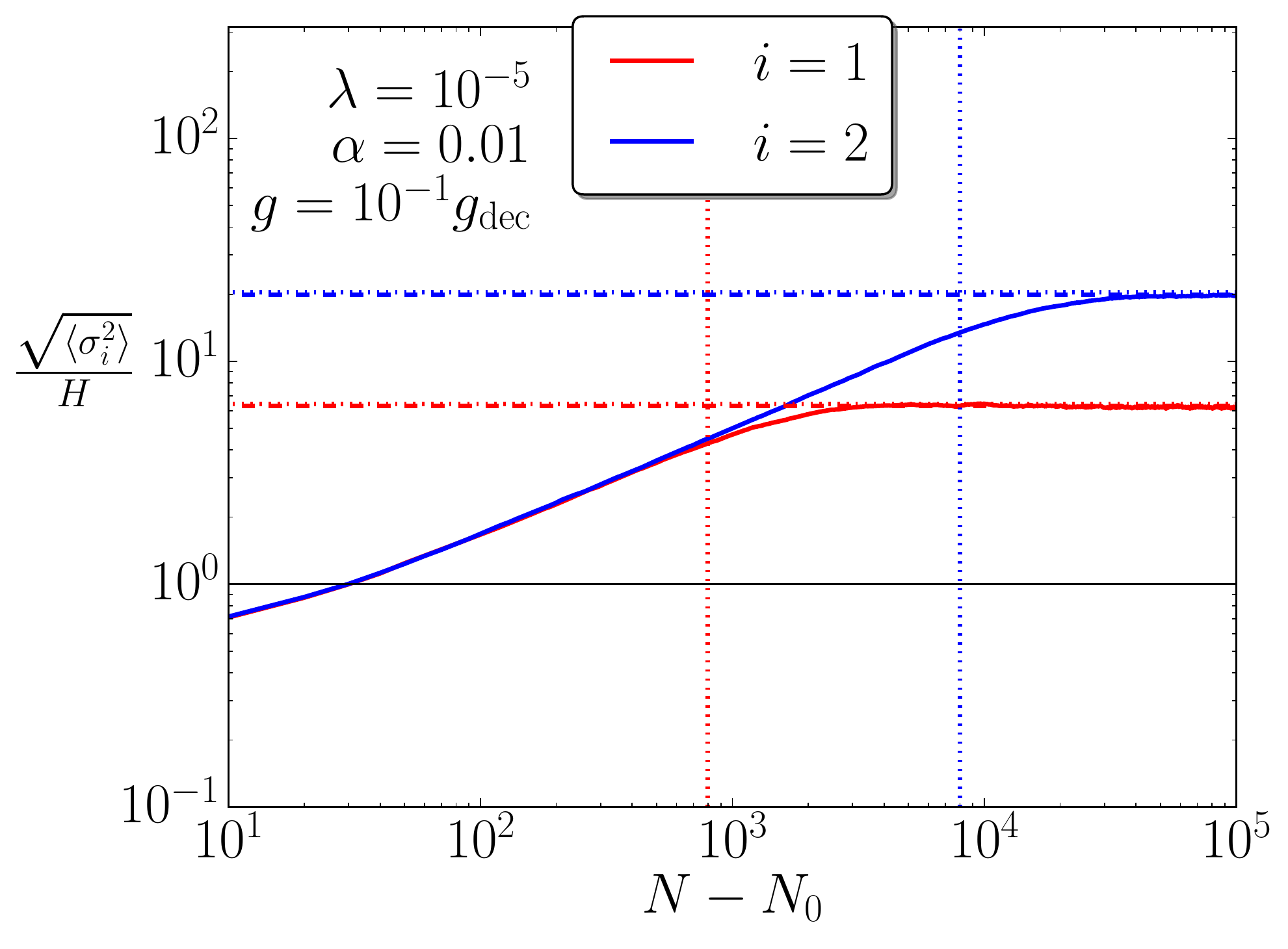}
\includegraphics[width=0.45\textwidth]{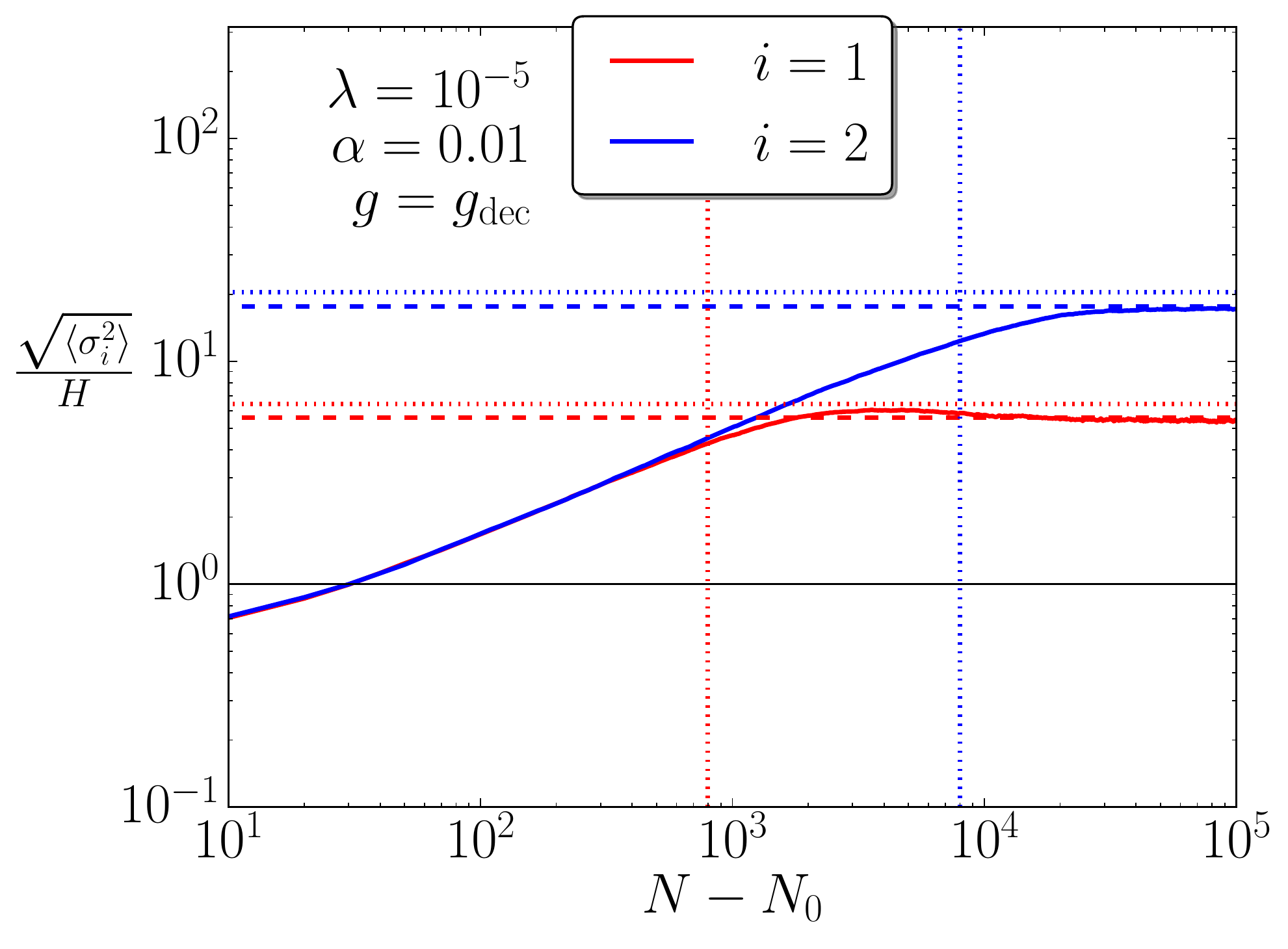} \\
\includegraphics[width=0.45\textwidth]{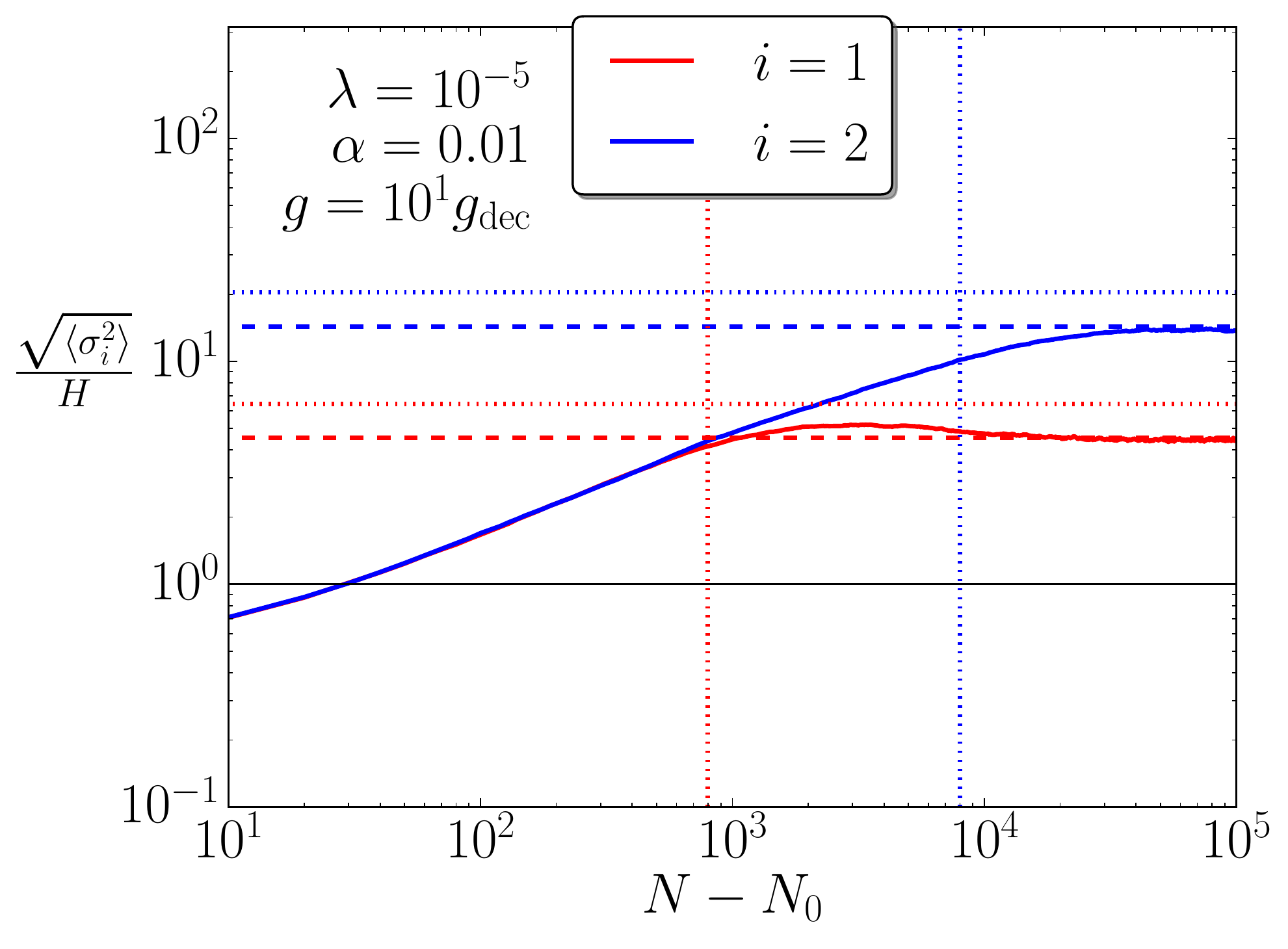}
\includegraphics[width=0.45\textwidth]{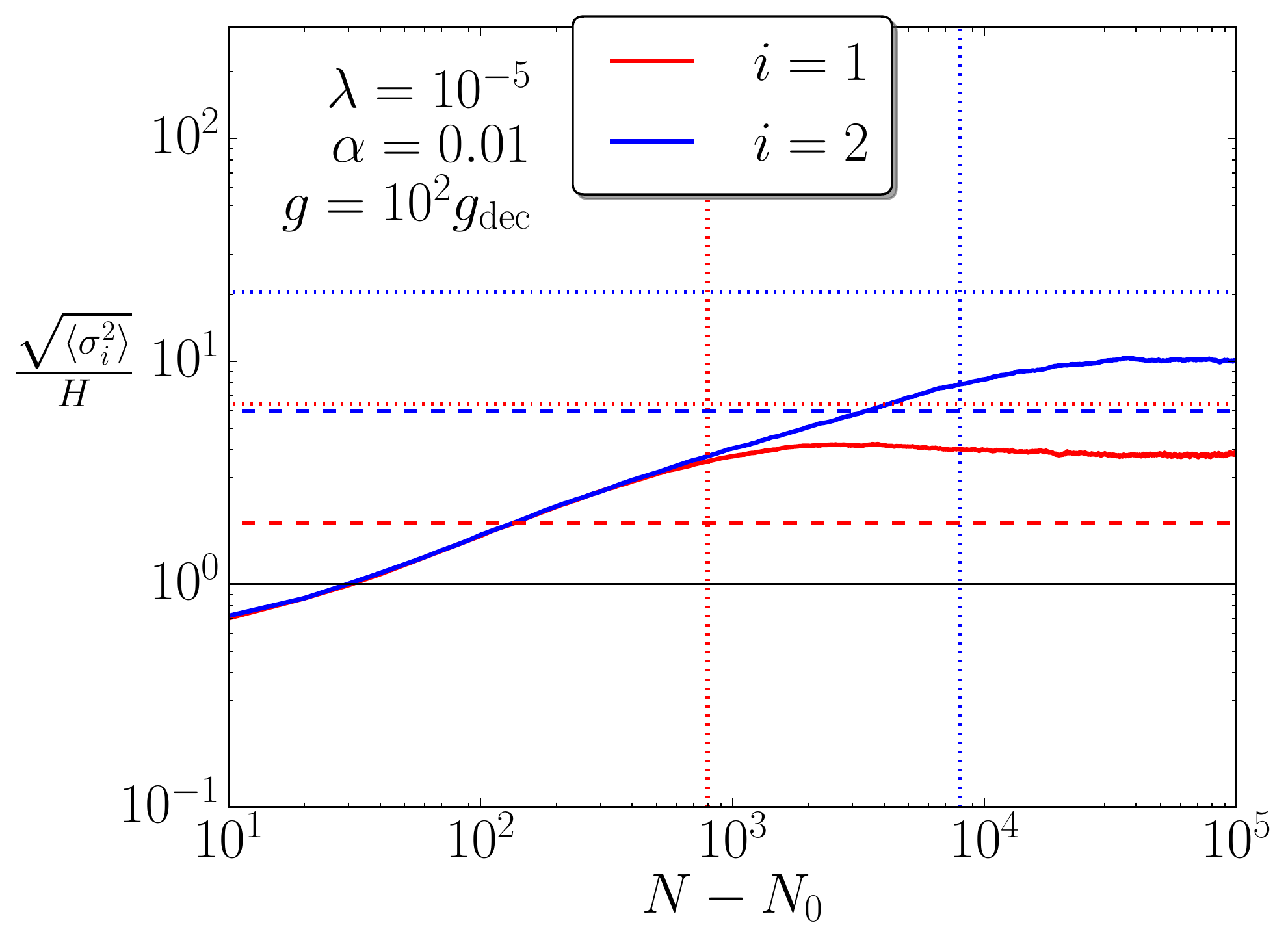}
\caption{~\label{fig:quartic-plots} The numerically evaluated (solid lines) time evolution of the variance for each spectator field in the case of the $V_{\rm C}$ potential (see \Eq{eq:VCpot}), where variance of both fields is initialised at $\left\langle \sigma_i^2\right\rangle = 0$ and $10^4$ realisations of \Eq{eq:langevin} were used in \href{https://github.com/umbralcalc/nfield}{\texttt{nfield}}. Dotted horizontal and dotted vertical lines represent the stationary variances (Eqs. \eqref{eq:stationary_VC_variances_1} and \eqref{eq:stationary_VC_variances_2}) and equilibration timescales (Eqs. \eqref{eq:stationary_VC_Neq_1} and \eqref{eq:stationary_VC_Neq_2}), respectively, computed in the decoupled limit $g\leq g_{\rm dec}$. The dashed horizontal lines use an alternative method to derive the stationary variance by numerically evaluating the second moment of \Eq{eq:exp-stat-dist}. The number of realisations used in each case is $10^4$. }
\end{center}
\end{figure}

\begin{figure}[t]
\begin{center}
\includegraphics[width=0.45\textwidth]{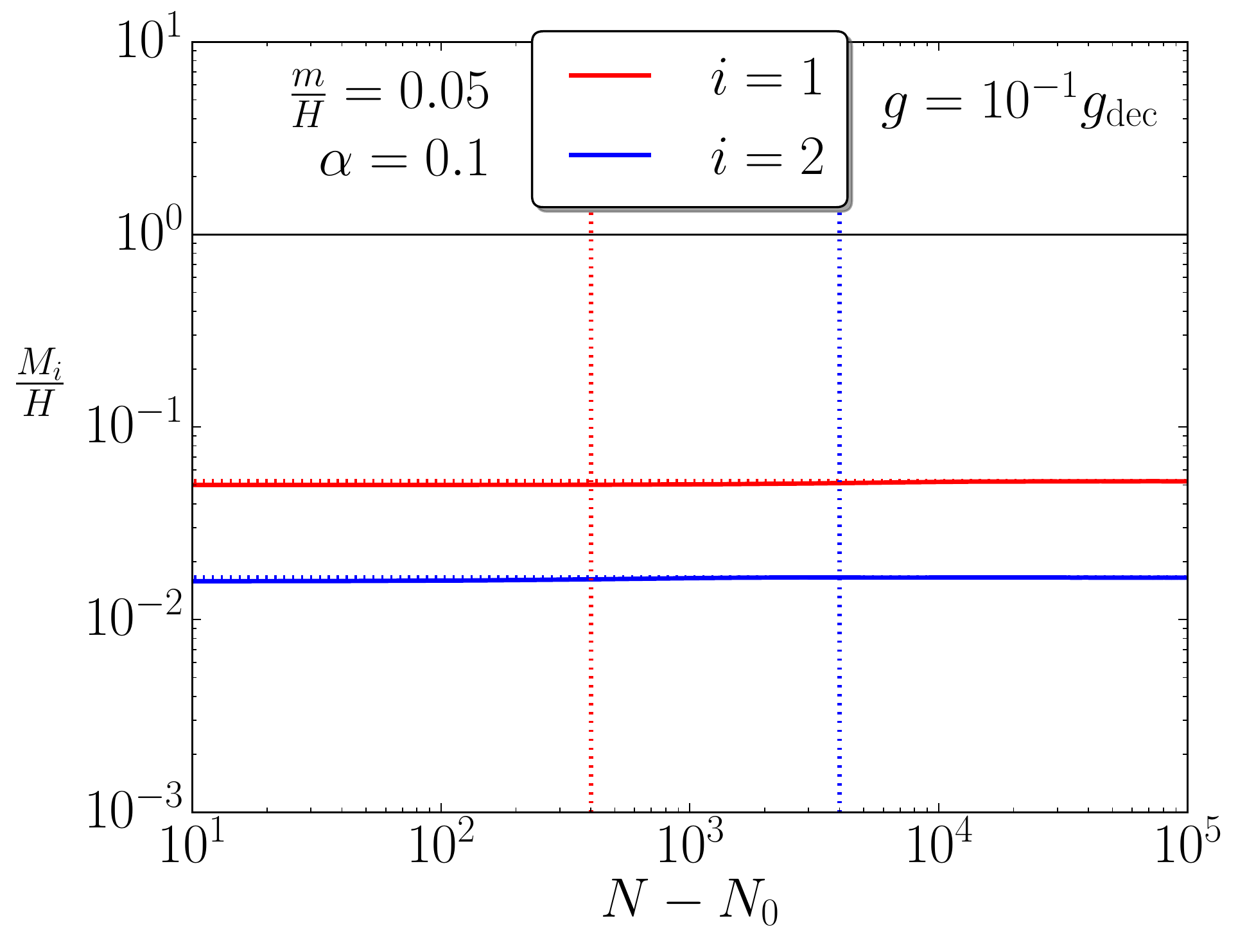}
\includegraphics[width=0.45\textwidth]{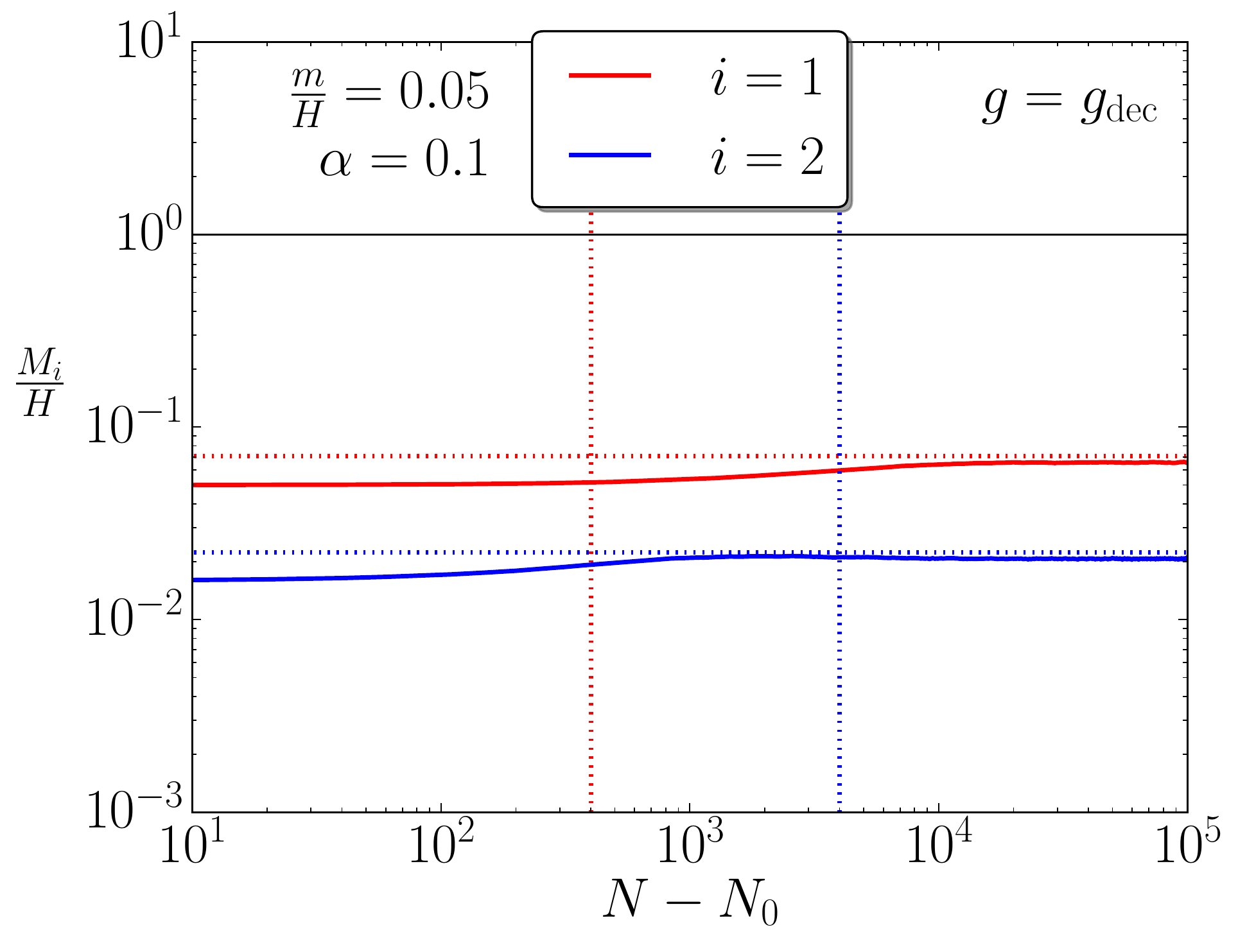} \\
\includegraphics[width=0.45\textwidth]{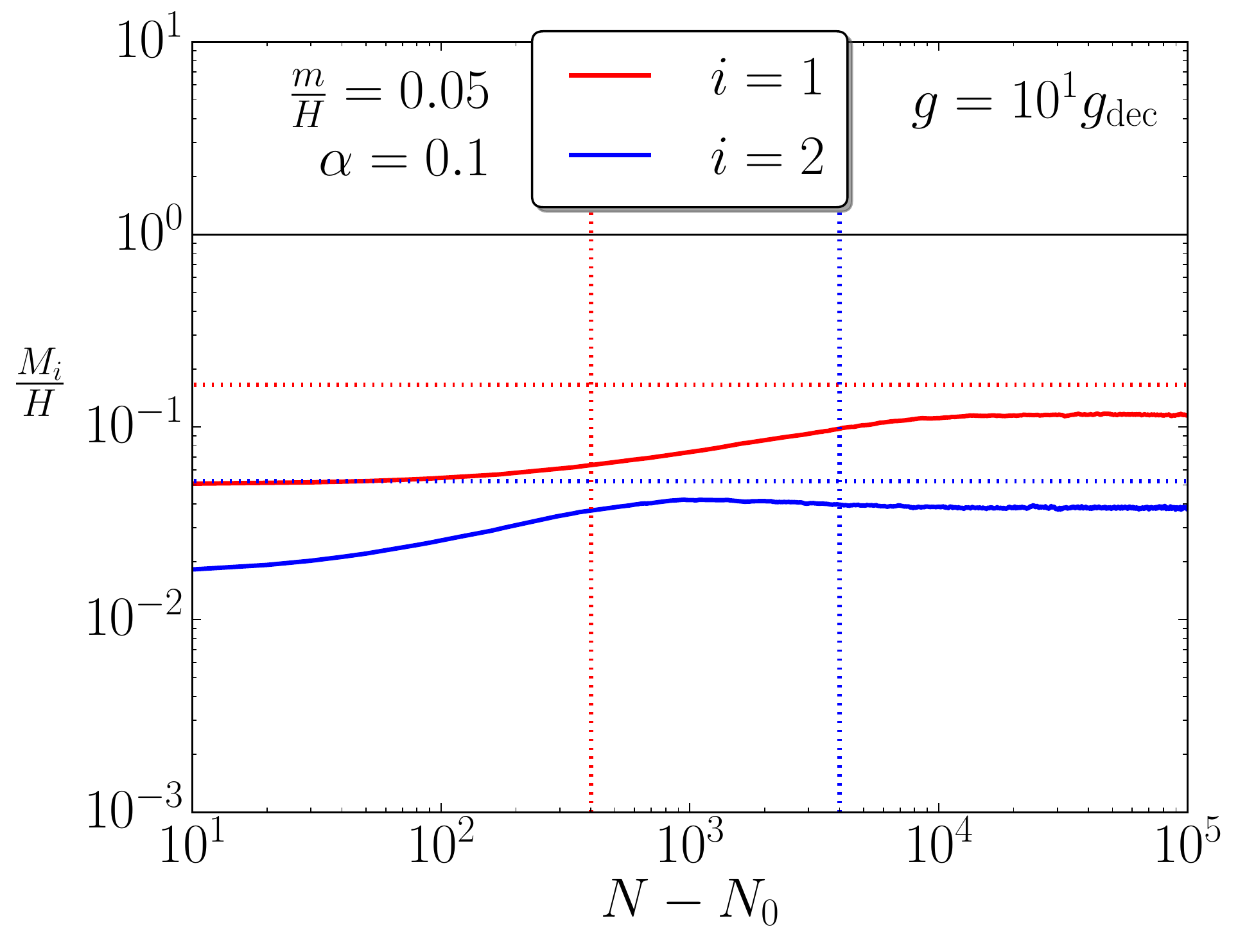}
\includegraphics[width=0.45\textwidth]{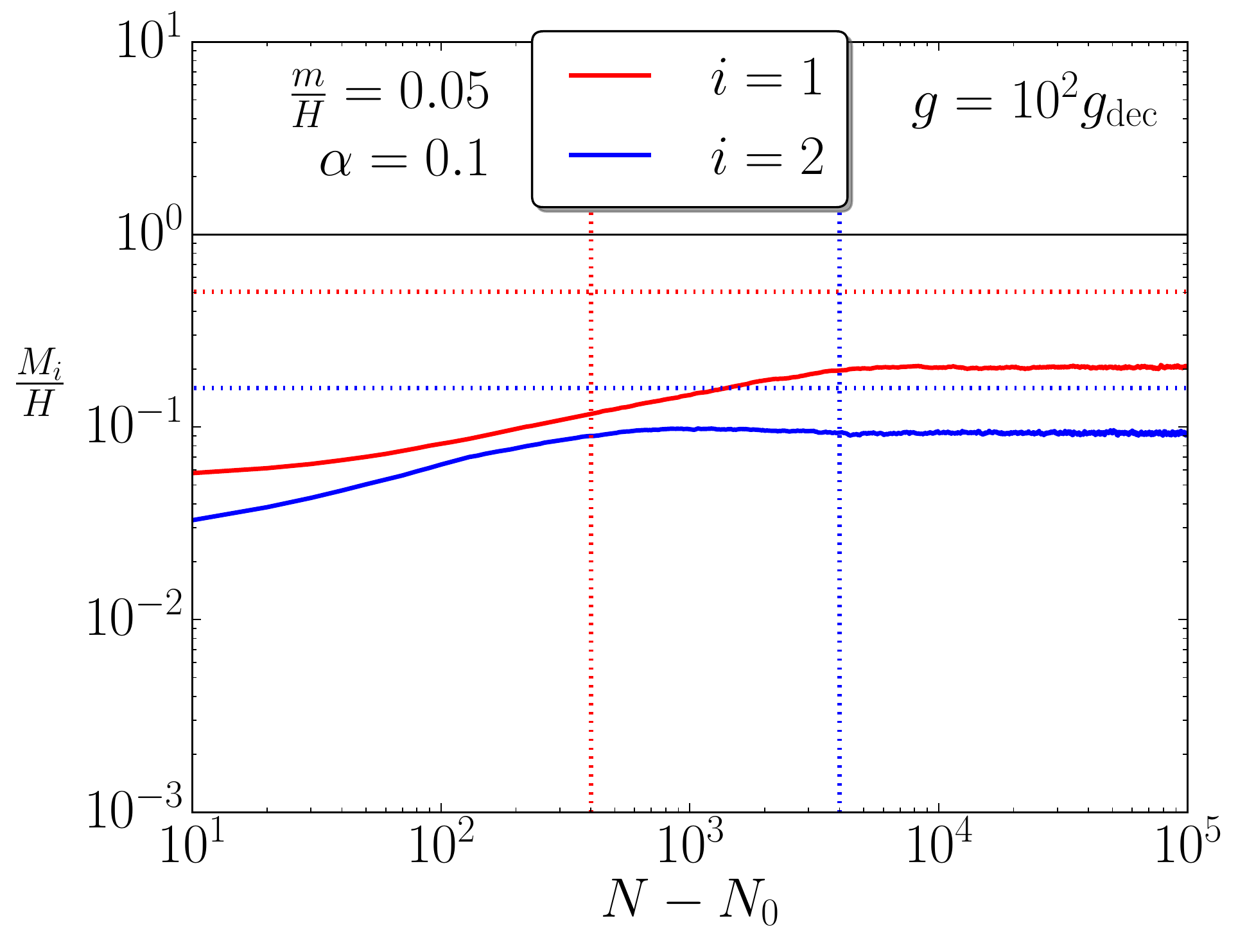}
\caption{~\label{fig:meff-quadratic-plots} The numerically evaluated (solid lines) time evolution of the effective mass (see \Eq{eq:effm}) for each spectator field in \Fig{fig:quadratic-plots}. Dotted horizontal and dotted vertical lines represent $M_i$ derived using the stationary variances (Eqs. \eqref{eq:stationary_VB_variances_1} and \eqref{eq:stationary_VB_variances_2}) and the equilibration timescales (Eqs. \eqref{eq:stationary_VB_Neq_1} and \eqref{eq:stationary_VB_Neq_2}), respectively. }
\end{center}
\end{figure}

\begin{figure}[t]
\begin{center}
\includegraphics[width=0.45\textwidth]{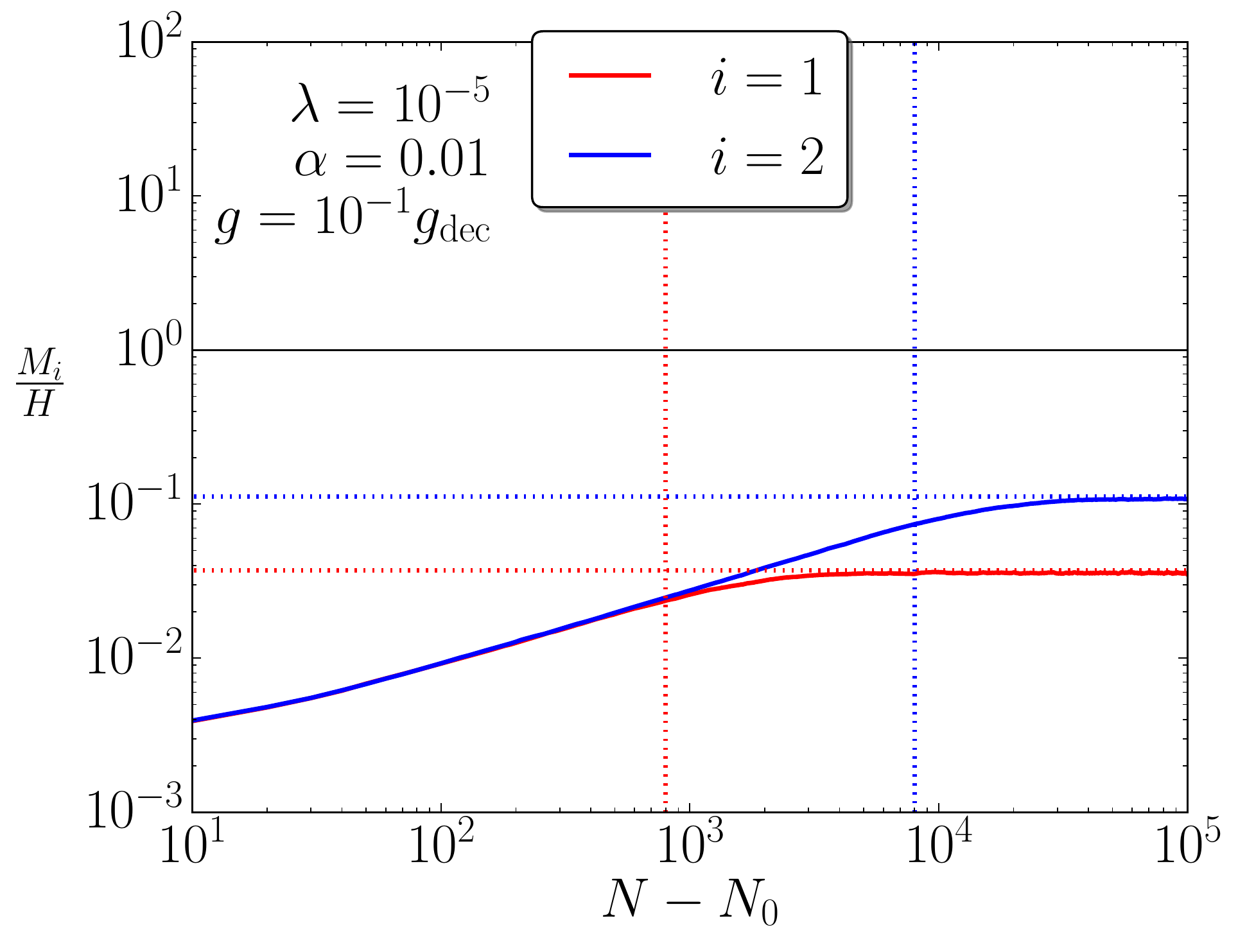}
\includegraphics[width=0.45\textwidth]{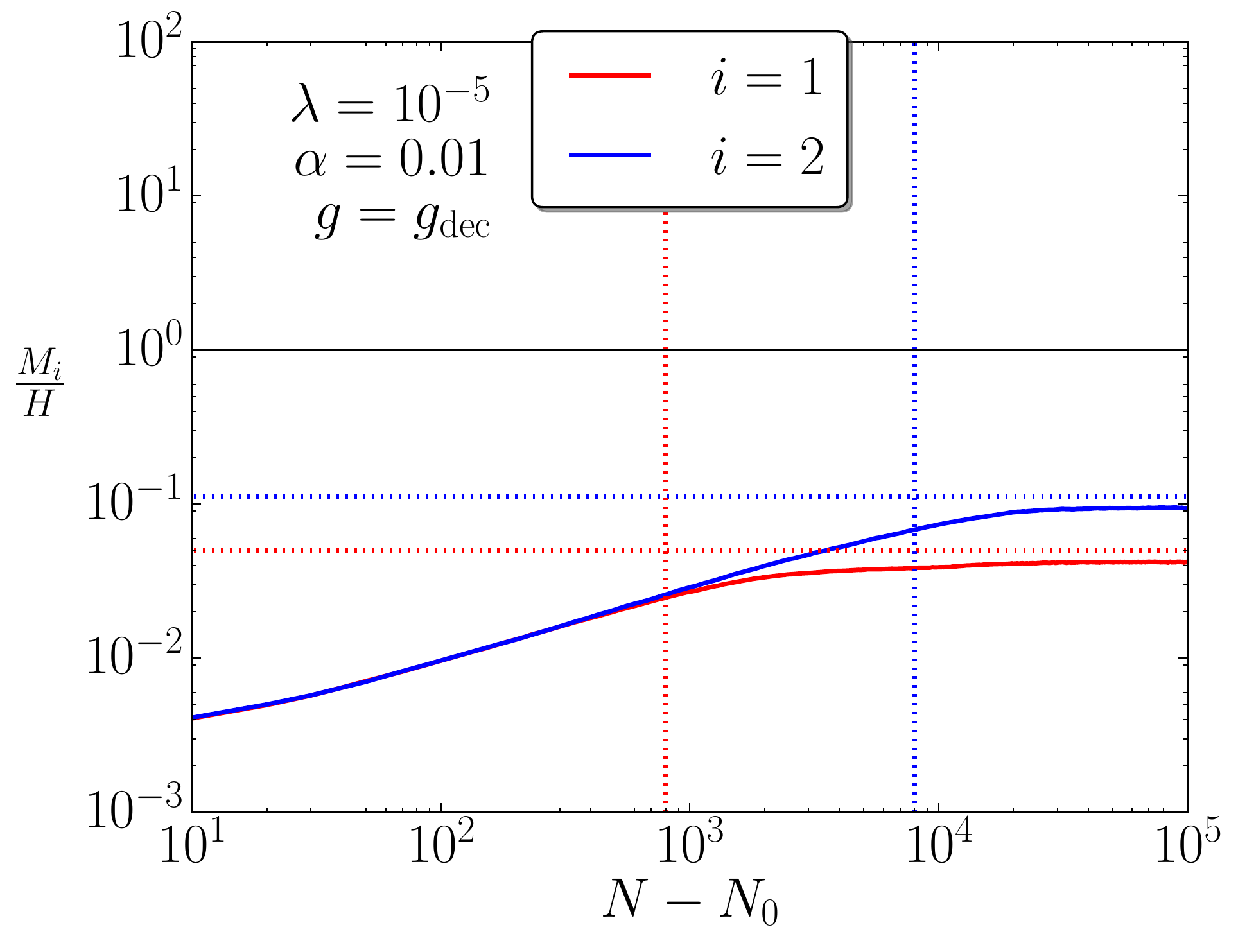} \\
\includegraphics[width=0.45\textwidth]{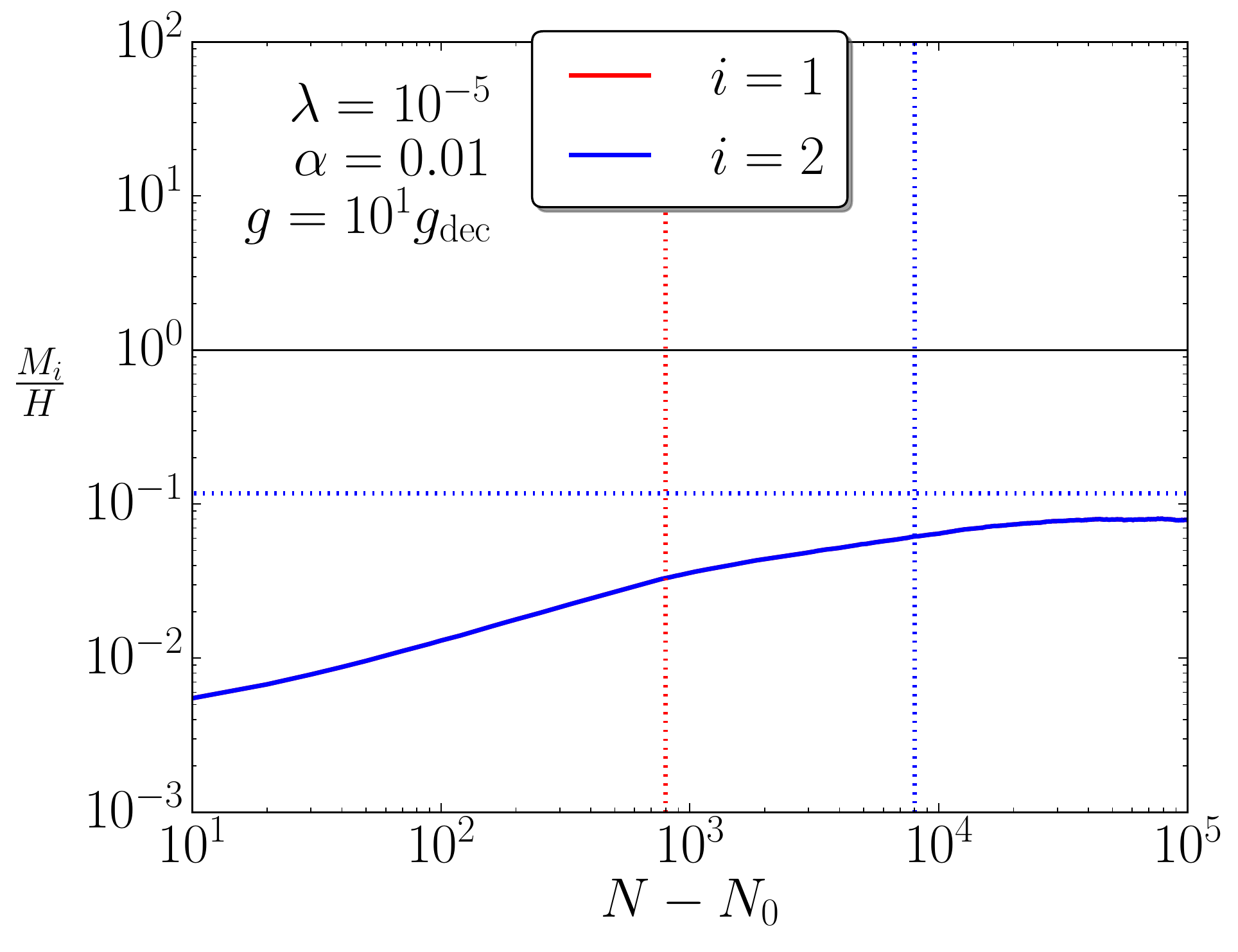}
\includegraphics[width=0.45\textwidth]{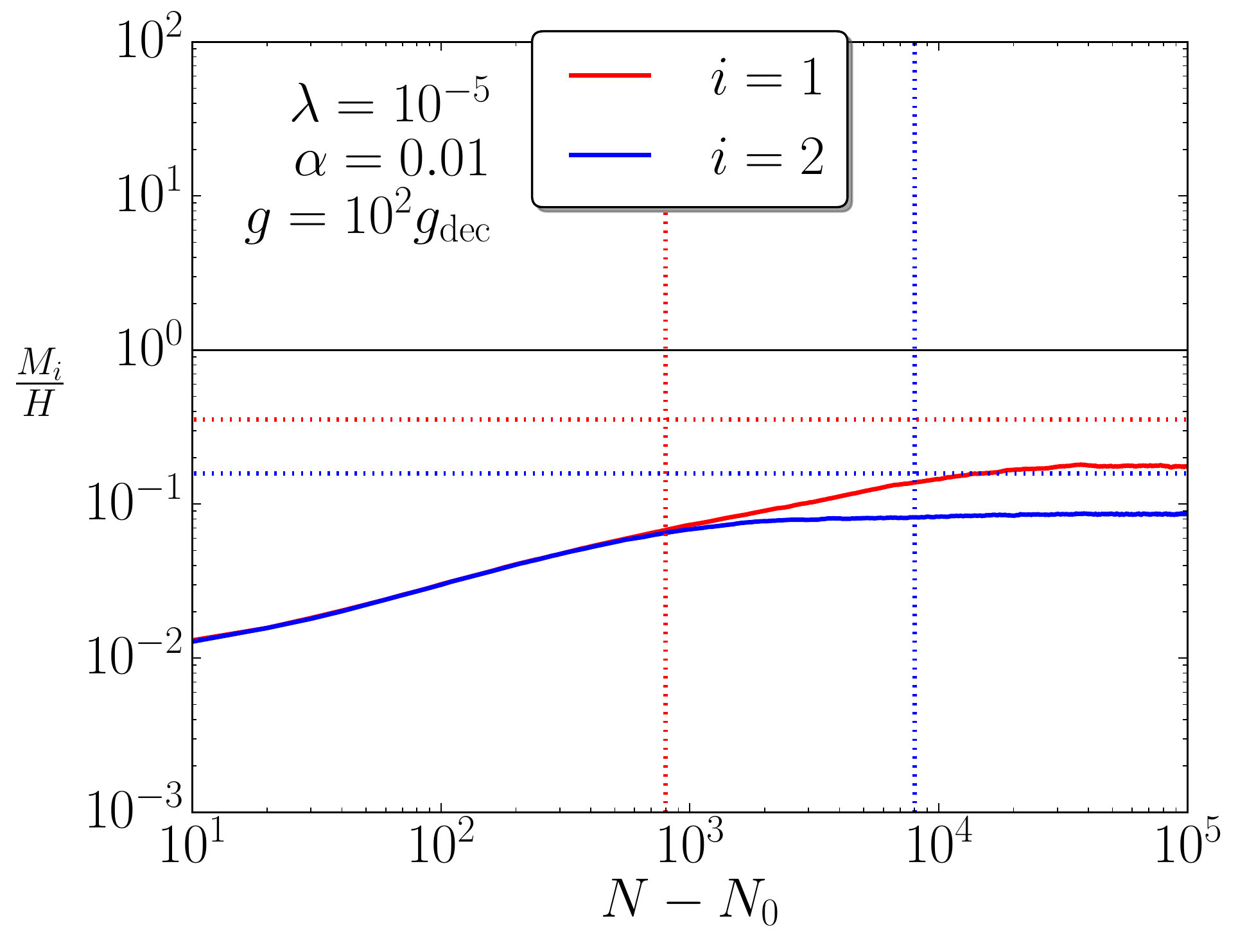}
\caption{~\label{fig:meff-quartic-plots} The numerically evaluated (solid lines) time evolution of the effective mass (see \Eq{eq:effm}) for each spectator field in \Fig{fig:quartic-plots}. Dotted horizontal and dotted vertical lines represent $M_i$ derived using the stationary variances (Eqs. \eqref{eq:stationary_VC_variances_1} and \eqref{eq:stationary_VC_variances_2}) and the equilibration timescales (Eqs. \eqref{eq:stationary_VC_Neq_1} and \eqref{eq:stationary_VC_Neq_2}), respectively. }
\end{center}
\end{figure}

If $g > g_{\rm dec}$ and $\alpha \neq 1$ however, then neither the equations above, nor the symmetry of $P(\sigma_i,N)$, can be exploited for analytic calculations and hence one must rely upon the numerically evaluated solution in order to study the system. In \Fig{fig:quadratic-plots} and \Fig{fig:quartic-plots} we plot these numerical solutions (and their corresponding effective masses in \Fig{fig:meff-quadratic-plots} and \Fig{fig:meff-quartic-plots}) given some specific values of $( m \, {\rm or} \, \lambda , \alpha , g )$ for both fields in the quadratic and quartic potentials, respectively. The variances are all initialised with $\left\langle \sigma_i^2\right\rangle = 0$ and hence the number of \efold{s} it takes for each solution to reach the effectively decoupled stationary values (dotted horizontal lines in the relevant colour using Eqs. \eqref{eq:stationary_VB_variances_1}, \eqref{eq:stationary_VB_variances_2}, \eqref{eq:stationary_VC_variances_1} and \eqref{eq:stationary_VC_variances_2}) is well-approximated by the analytic relaxation timescales derived in Eqs. \eqref{eq:stationary_VB_Neq_1}, \eqref{eq:stationary_VB_Neq_2}, \eqref{eq:stationary_VC_Neq_1} and \eqref{eq:stationary_VC_Neq_2} (depicted with vertical dotted lines in the relevant colour) in cases where $g\leq g_{\rm dec}$ (the top row plots of both sets of Figs.). In all plots, one can also clearly see the strong deviation from the decoupled predictions with larger values of $g$, which highlights the importance for a numerical solution from \href{https://github.com/umbralcalc/nfield}{\texttt{nfield}} in this large regime of parameter values to obtain the correct equilibrium as well as out-of-equilibrium behaviour.

\begin{figure}[t]
\begin{center}
\includegraphics[width=0.65\textwidth]{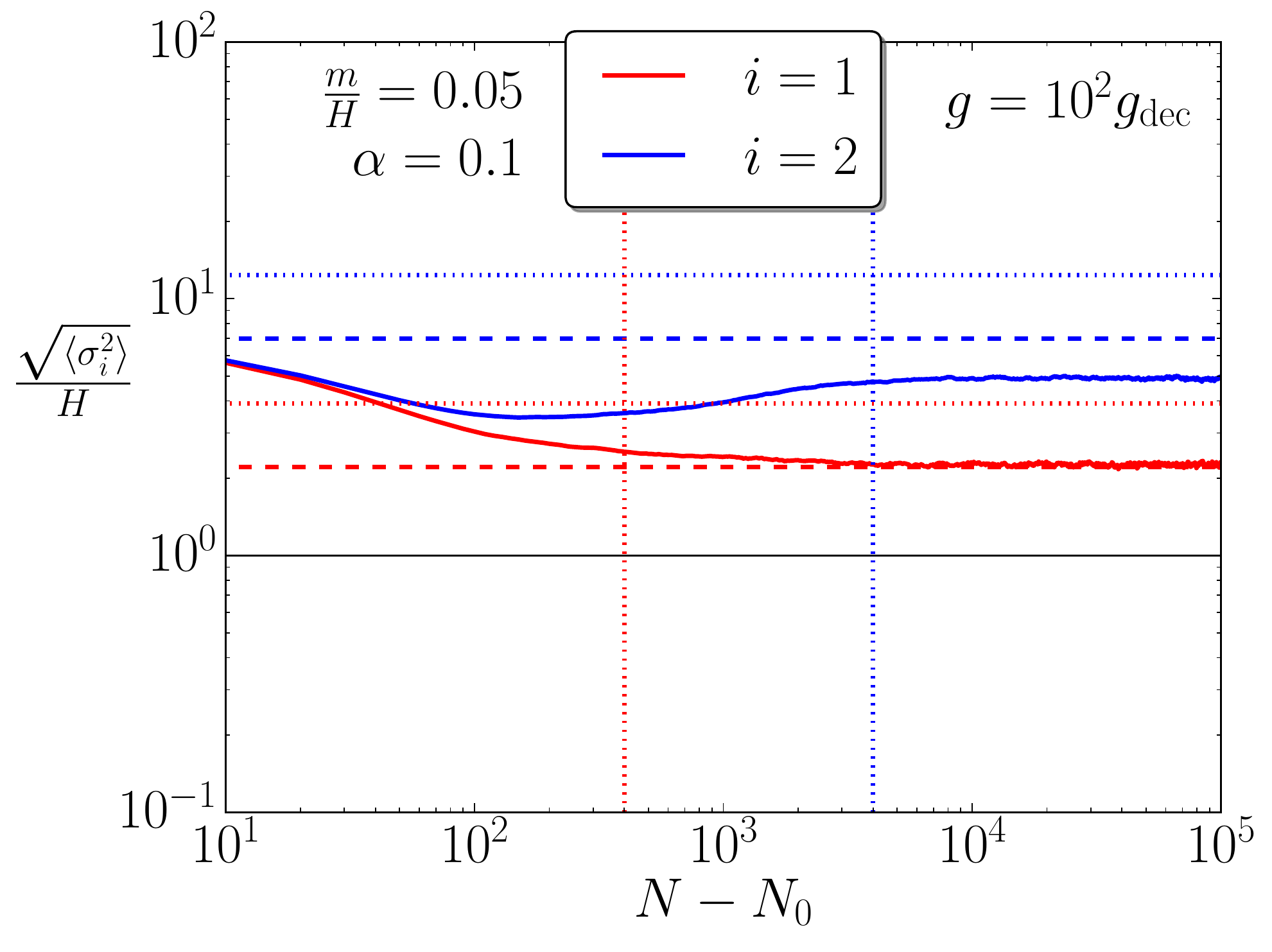}
\caption{~\label{fig:varyinitcond-VB-variance} An illustrative re-plotting of the numerical variance evolution, with an initial condition much closer to the analytic stationary variance derived from the second moment of \Eq{eq:exp-stat-dist}, for the case in the bottom right-hand corner of \Fig{fig:quadratic-plots}.  }
\end{center}
\end{figure}

\subsection{Non-vanishing probability currents} \label{sec:non-vanish}

In \Fig{fig:quadratic-plots} there is also an important anomaly which appears to be repeated in \Fig{fig:quartic-plots}. In both figures we have also provided (dashed horizontal lines) an alternative calculation for the stationary variance using the numerically calculated second moment of \Eq{eq:exp-stat-dist}. There is generally excellent agreement  between this solution and the one obtained from the many realisations of \Eq{eq:langevin} in \href{https://github.com/umbralcalc/nfield}{\texttt{nfield}} for $g\leq 10g_{\rm dec}$, however these no longer agree precisely when $g = 10^2g_{\rm dec}$ in both sets of plots. This deviation has been checked for numerical robustness by increasing the number of Langevin realisations to $10^5$ and altering the initial conditions --- see \Fig{fig:varyinitcond-VB-variance} for illustration.

\begin{figure}[t]
\begin{center}
\includegraphics[width=0.60\textwidth]{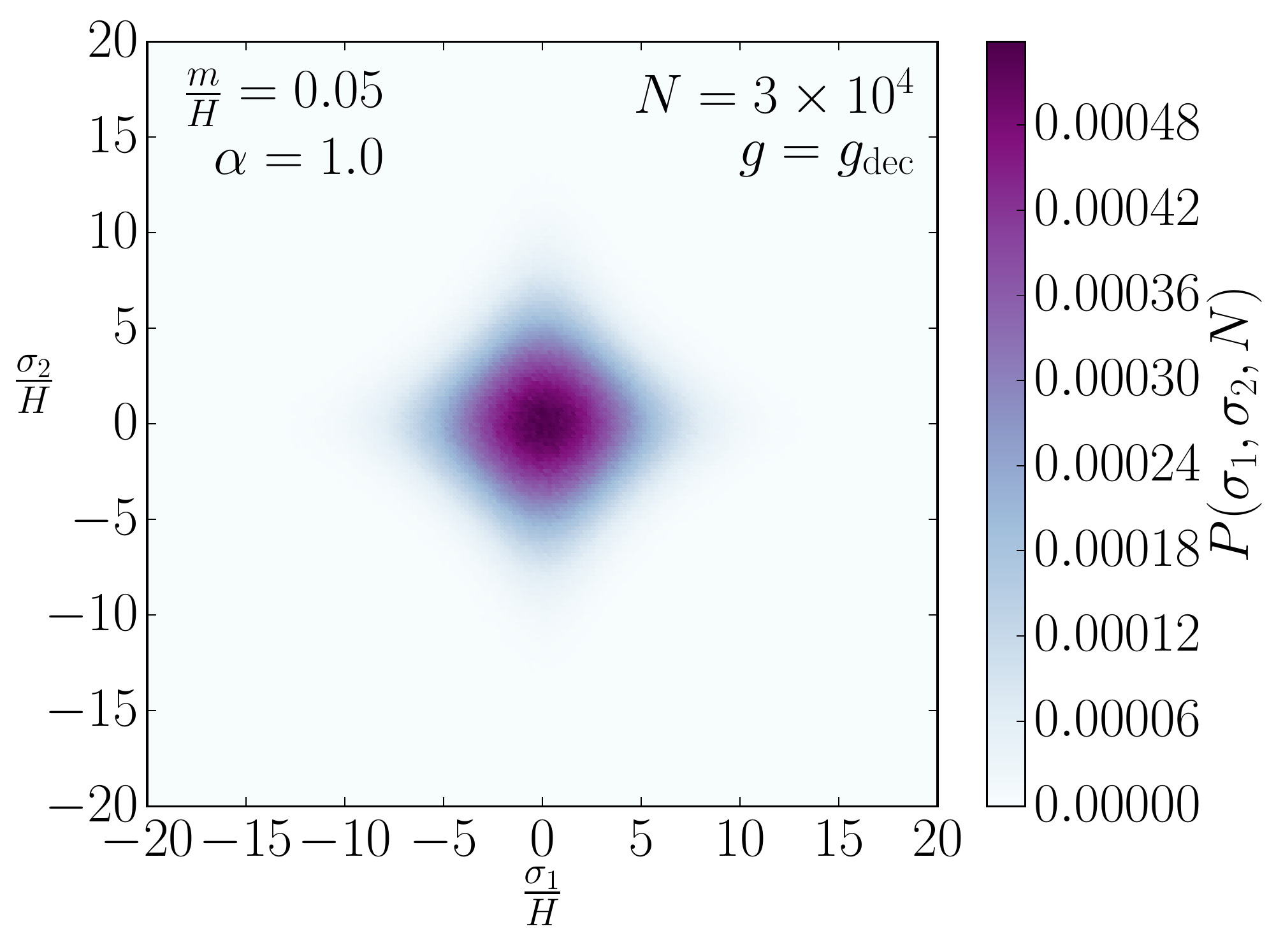} \\
\includegraphics[width=0.45\textwidth]{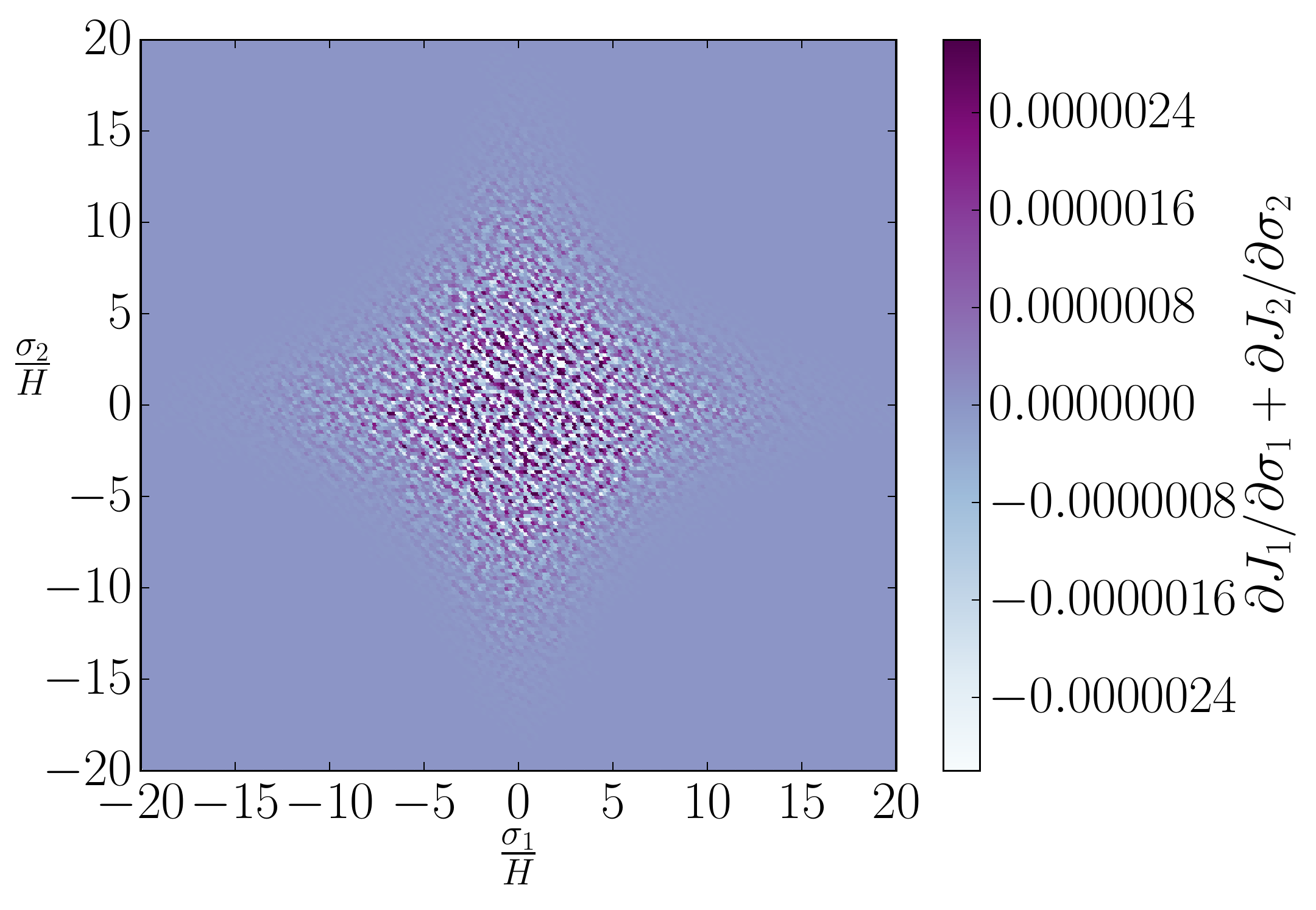}  \includegraphics[width=0.45\textwidth]{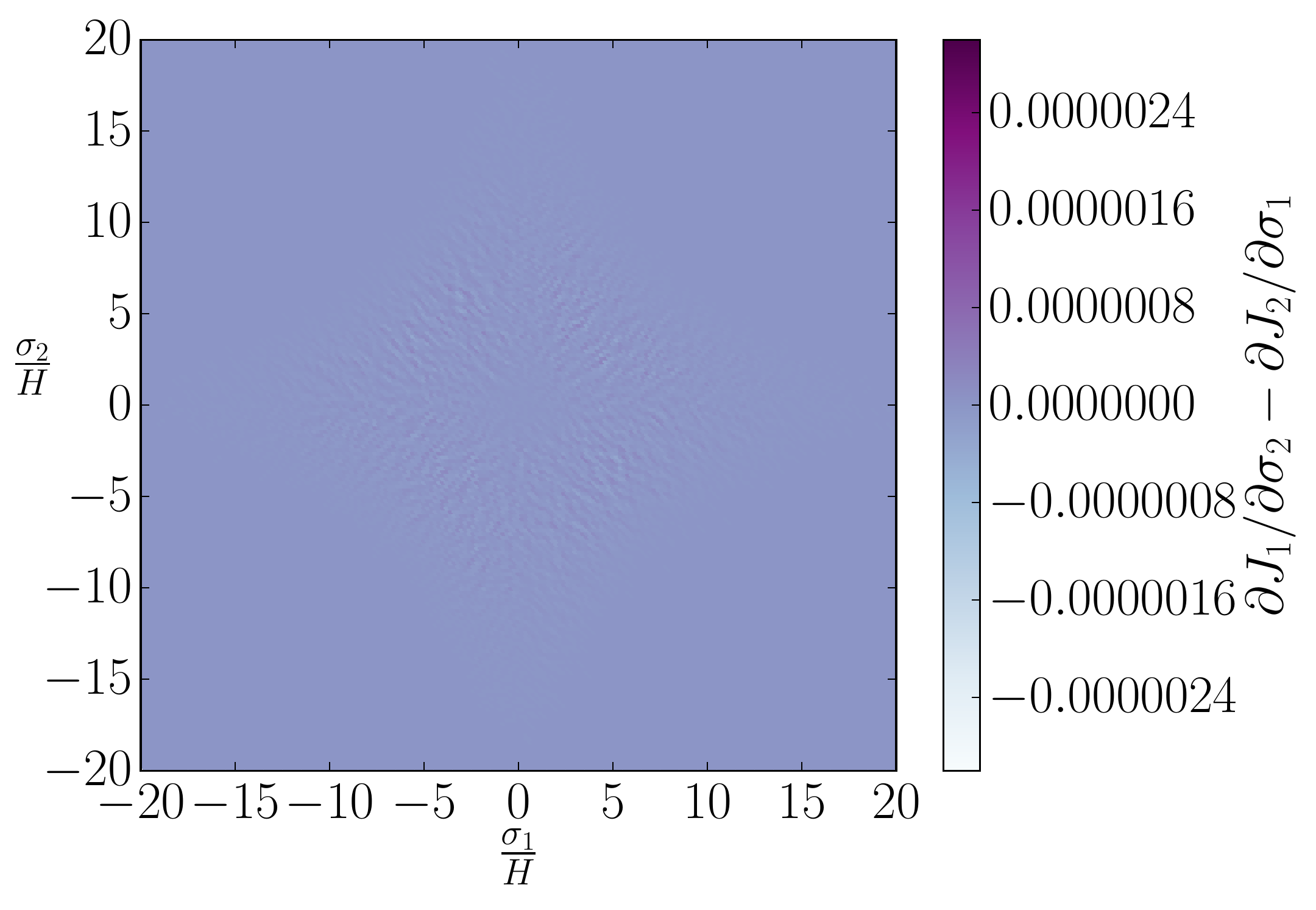}
\caption{~\label{fig:pdf-cur-plot1} Using specific parameter choices of the $V_{\rm B}$ potential (see \Eq{eq:VBpot}) we plot the binned stationary probability density $P_{\rm stat}$ (on top), probability current divergence $\nabla \cdot \boldsymbol{J}$ (bottom left) and the only non-zero component of the probability current curl $\nabla \times \boldsymbol{J}$ (bottom right). These have all been numerically obtained from $10^7$ realisations of \Eq{eq:langevin} in the \href{https://github.com/umbralcalc/nfield}{\texttt{nfield}} code. }
\end{center}
\end{figure}

\begin{figure}[t]
\begin{center}
\includegraphics[width=0.45\textwidth]{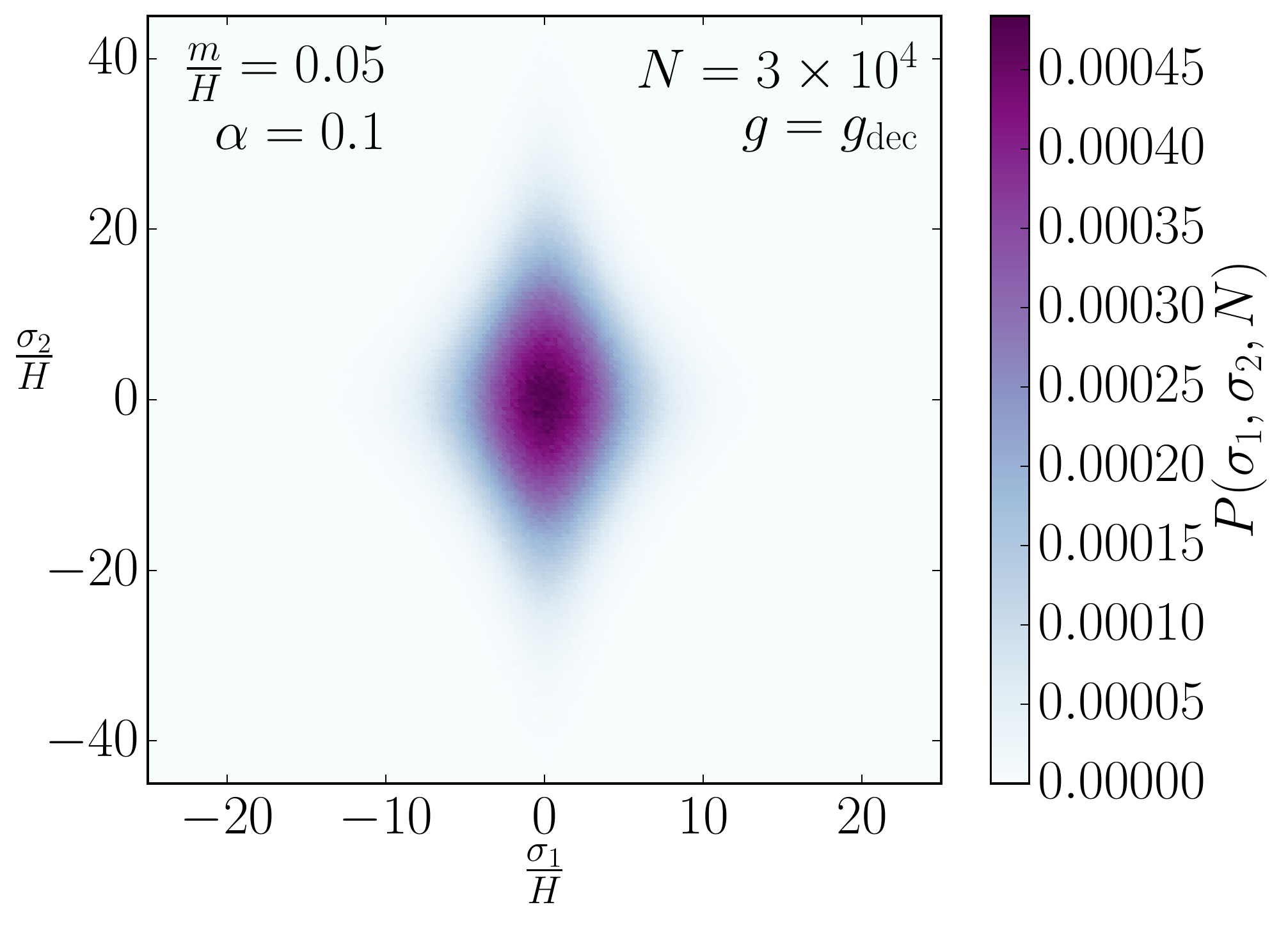}
\includegraphics[width=0.45\textwidth]{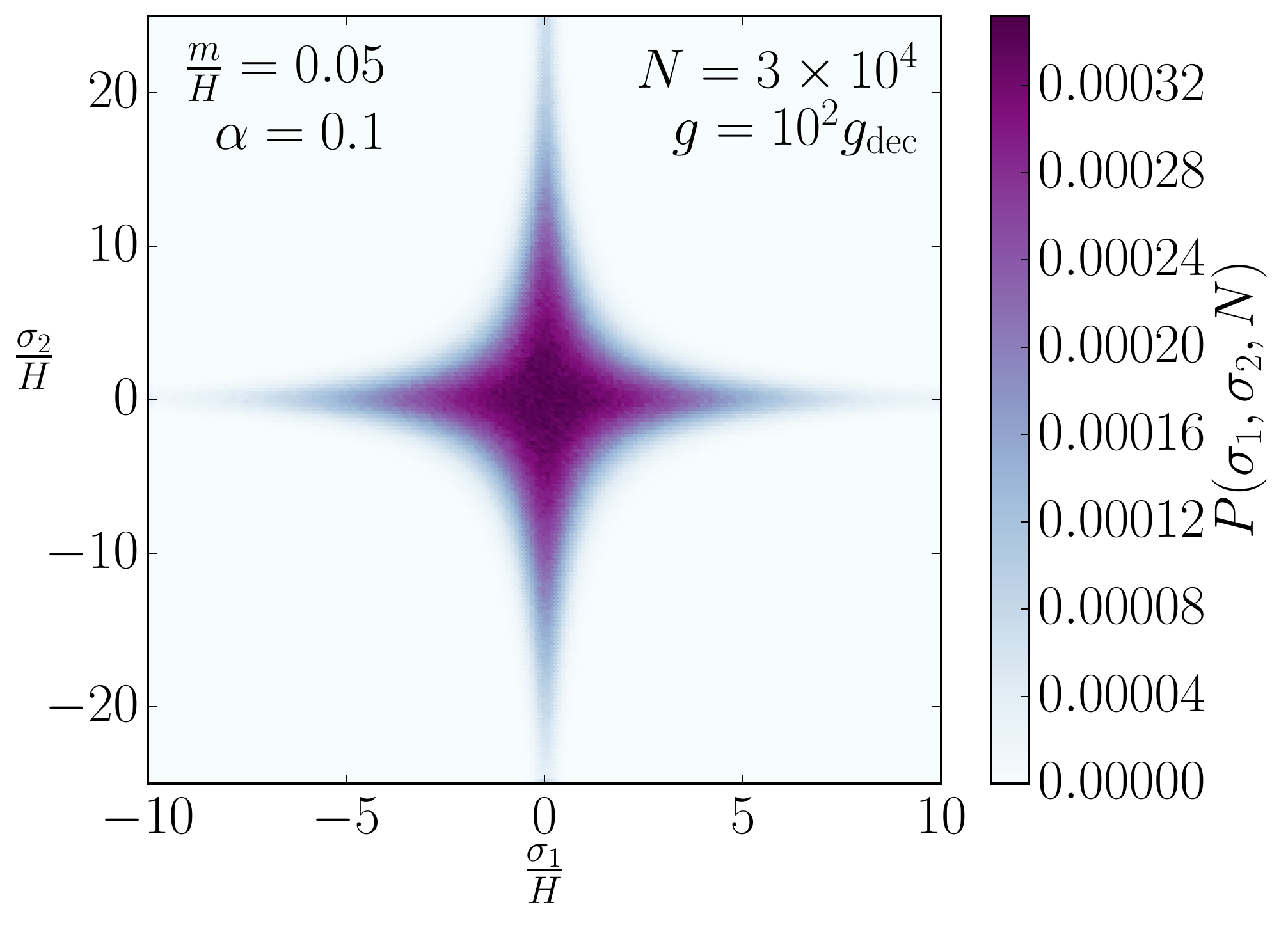}  \\
\includegraphics[width=0.45\textwidth]{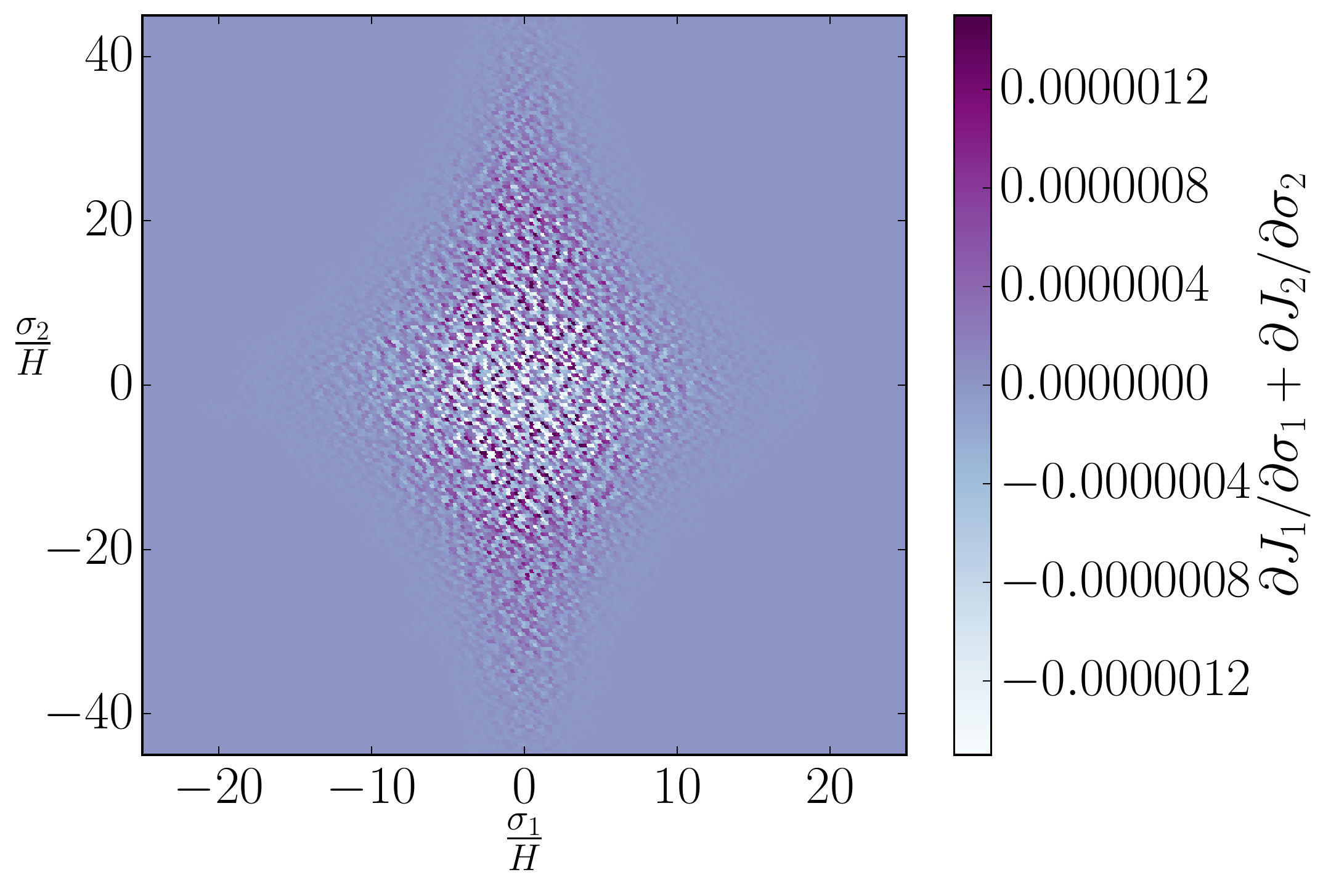}
\includegraphics[width=0.45\textwidth]{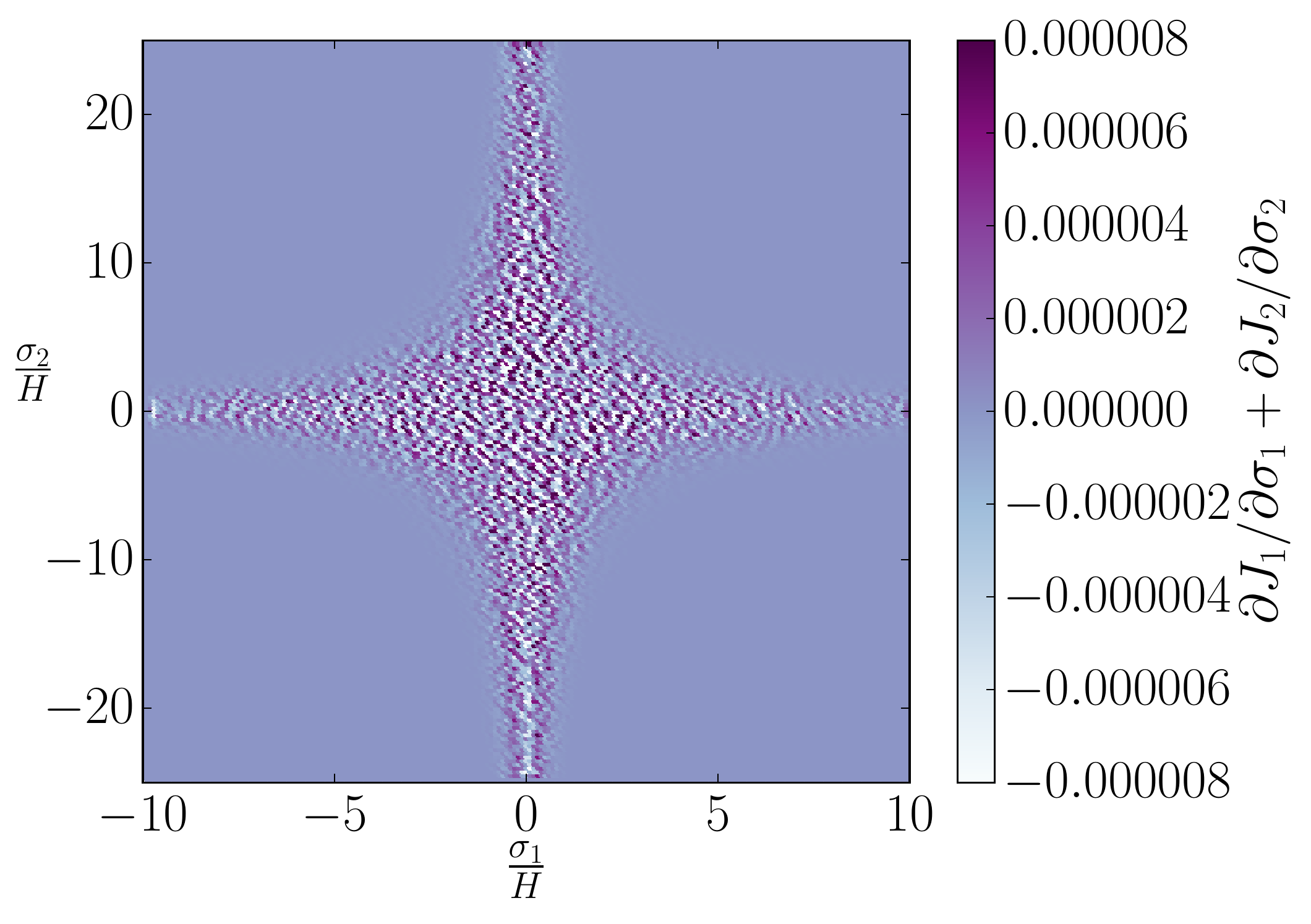} \\
\includegraphics[width=0.45\textwidth]{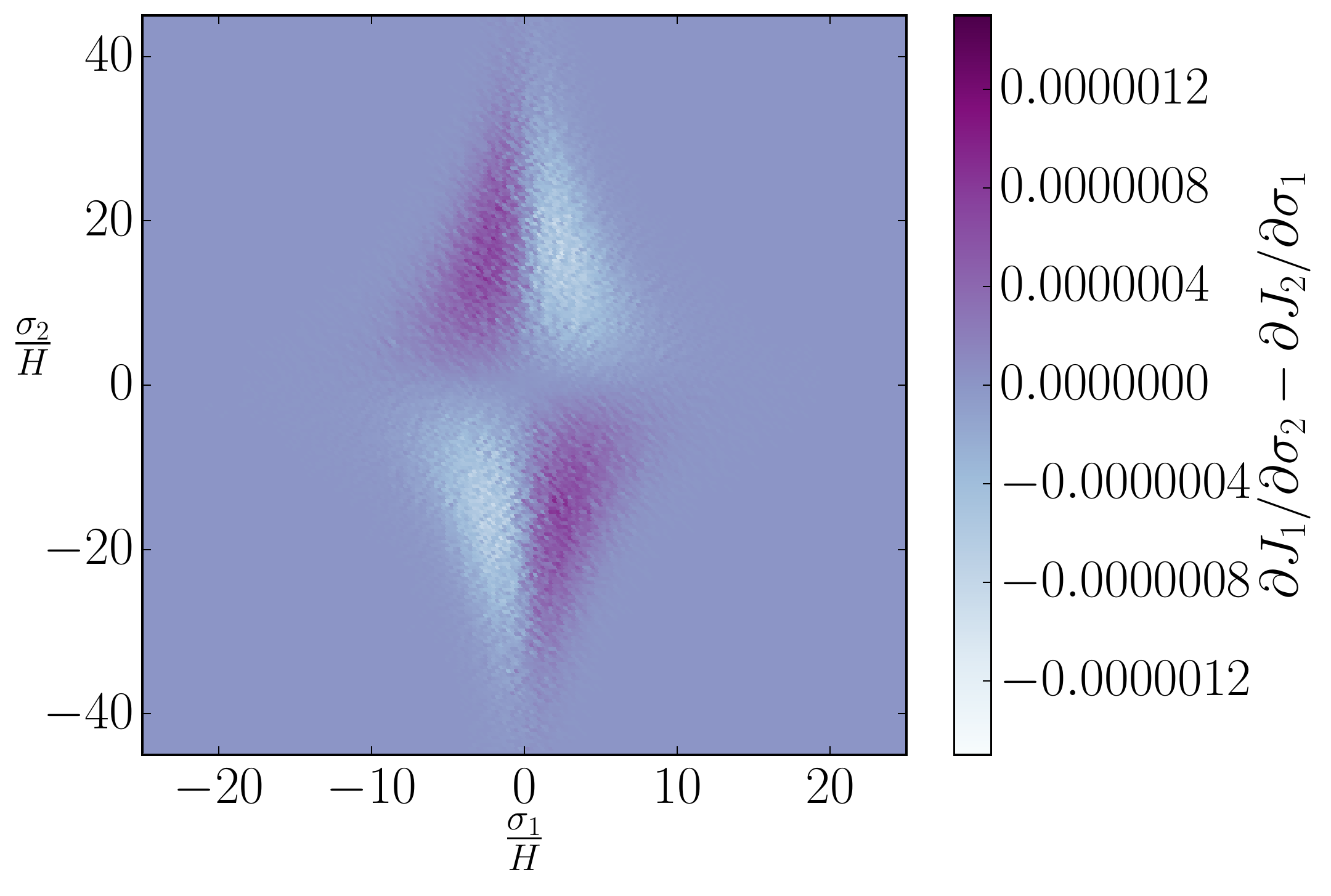}
\includegraphics[width=0.45\textwidth]{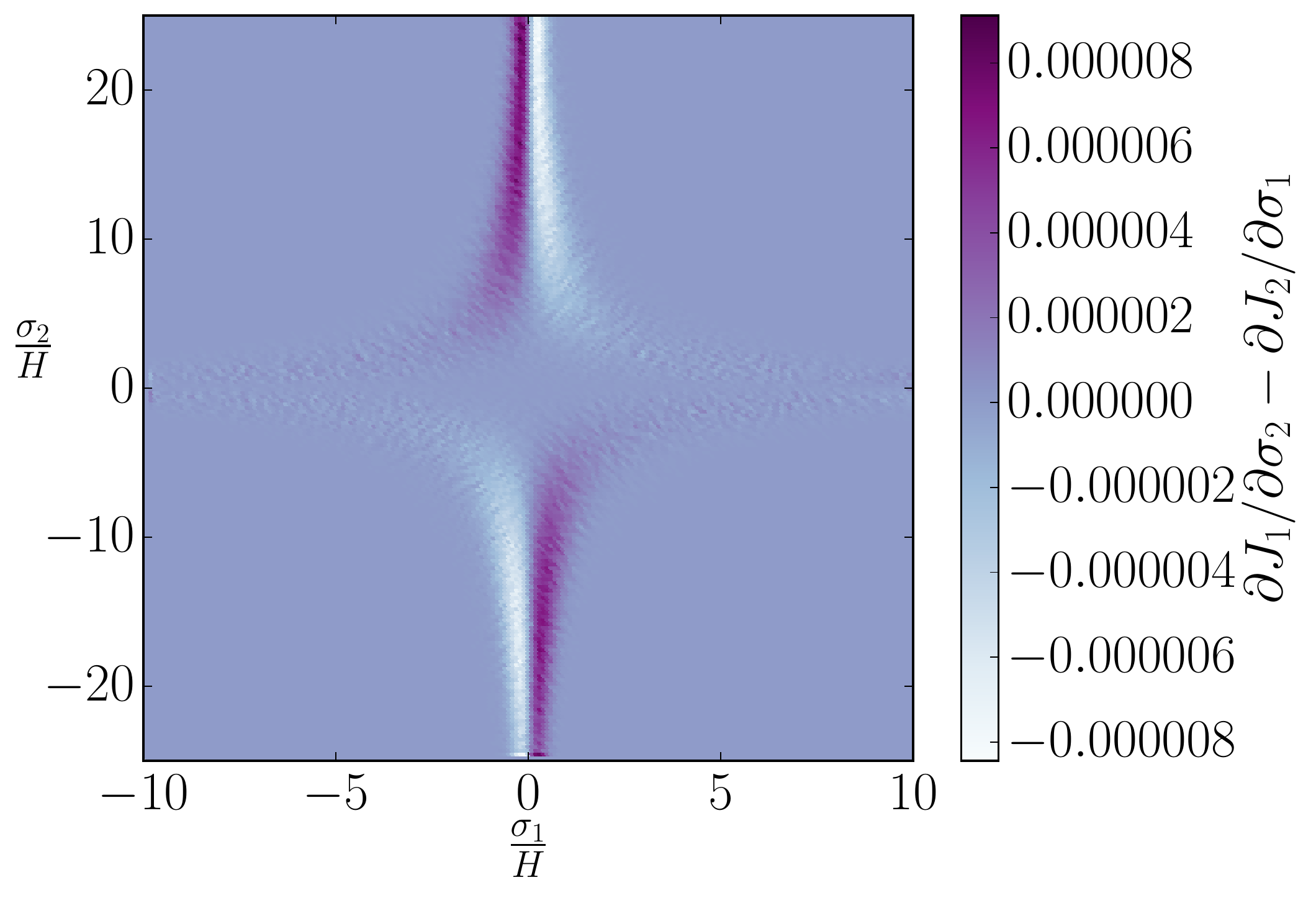}
\caption{~\label{fig:pdf-cur-plot2} Using specific parameter choices of the $V_{\rm B}$ potential (see \Eq{eq:VBpot}) we plot the binned stationary probability density $P_{\rm stat}$ (top row), probability current divergence $\nabla \cdot \boldsymbol{J}$ (middle row) and the only non-zero component of the probability current curl $\nabla \times \boldsymbol{J}$ (bottom row). These have all been numerically obtained from $10^7$ realisations of \Eq{eq:langevin} in the \href{https://github.com/umbralcalc/nfield}{\texttt{nfield}} code. }
\end{center}
\end{figure}

We are left with the interesting conclusion that for a sufficiently large coupling $g$, and an asymmetric potential induced by the mass hierarchy parameter $\alpha < 1$, \Eq{eq:exp-stat-dist} is no longer sufficient to describe the stationary probability distribution. In \Sec{sec:proof-symmetric} we conjectured that \Eq{eq:exp-stat-dist} is the stationary solution for symmetric potentials (here when $\alpha =1$). However, when $\alpha <1$, since only the divergence of the probability current $\nabla \cdot \boldsymbol{J}$ must vanish and \emph{not} its curl $\nabla \times \boldsymbol{J}$, \Eq{eq:exp-stat-dist} is no longer the true stationary solution and hence the solution must be elucidated through full numerical evaluation of either \Eq{eq:langevin} or \Eq{eq:dist}. We have plotted $\nabla \times \boldsymbol{J}$ for different choices of parameter in \Fig{fig:pdf-cur-plot1} and \Fig{fig:pdf-cur-plot2}, where one can see in particular that the only component of $\nabla \times \boldsymbol{J}$ is much larger when $g$ is increased for the $\alpha =0.1$ cases plotted in \Fig{fig:pdf-cur-plot2}. As a further numerical check, we have verified that when the symmetry of the potential is restored ($\alpha = 1$) in \Fig{fig:pdf-cur-plot1}, the curl vanishes up to some numerical noise.

Note that $\nabla \times \boldsymbol{J}$ also vanishes at the origin in both \Fig{fig:pdf-cur-plot1} and \Fig{fig:pdf-cur-plot2}. This is confirmed by the analytic expression in \Eq{eq:curl-J}, in which the curl is indeed vanishing at the extrema $\partial V/\partial \sigma_i=\partial P/\partial \sigma_i=0$.

\section{Conclusions}
\label{sec:conclusions}

In this paper we have demonstrated the usefulness of numerical solutions in order to evaluate the variances of multiple light coupled fields during inflation. In doing so we have identified a lower limit $g_{\rm dec}$ in some example two-field potentials on the coupling $g$, for interactions of the form $\propto g \sigma_1^2\sigma_2^2$, below which the fields may be considered as effectively decoupled and the standard formulae for stationary variances may be used. We have further verified that for choices of $g\geq g_{\rm dec}$, the analytic decoupling approximation for the variances breaks down. In such situations, the solutions from either evaluating the moments of \Eq{eq:exp-stat-dist} (Eqs. \eqref{eq:VBpot} and \eqref{eq:VCpot}, stable in the stationary limit when the potential is either symmetric $\alpha =1$ or decoupled $g<g_{\rm dec}$) or full numerical solutions (for all potentials and generic initial conditions) are the methods to obtain correct values.

In \Sec{sec:proof-symmetric} we have given a general argument as to why it is possible for \Eq{eq:exp-stat-dist} to still remain stable for some symmetric potentials due to vanishing of the probability current everywhere in the domain. Conversely, we have shown that by breaking the symmetry in the potential (e.g. $\alpha \neq 1$ in Eqs. \eqref{eq:VBpot} and \eqref{eq:VCpot}) the form of \Eq{eq:exp-stat-dist} may no longer be stable as a solution to the stationary behaviour of the multi-spectator system. We have supported these conclusions with the numerically obtained figures provided in \Sec{sec:dec-coup} and Figs. \ref{fig:pdf-cur-plot1} and \ref{fig:pdf-cur-plot2}.

A simple generalisation for future work may be to check how this limit changes as the number of coupled fields $n_{\rm f}$ is increased, where we anticipate that because increasing $n_{\rm f}$ typically increases the contribution from the coupling to the effective mass of each field, the lower limit on $g=g_{\rm dec}$ should decrease in order to compensate. Due to the complexity of such a system, a numerical scheme such as the one we have developed in this paper\footnote{One can go to the following repository to access the code: \href{https://github.com/umbralcalc/nfield}{https://github.com/umbralcalc/nfield}.} (\href{https://github.com/umbralcalc/nfield}{\texttt{nfield}}) will likely be required for such an extension.

By considering an arbitrary $n_{\rm f}$ in the symmetric potential of \Eq{eq:VA} we have also discovered a critical value for $g=g_{\rm crit}$ that varies $\propto 1/n_{\rm f}$ (see \Eq{eq:gcrit-analytic} for a more precise form) above which the formation of stationary spectator condensates collapses to the Hubble rate. For values of $g>g_{\rm crit}$, we cannot yet precisely say that the formation of stationary condensates in such a potential is suppressed (as it is when increasing $g$ up to this point) because this phenomenon results from the effective mass $M_i$ of each field reaching ${\cal O}(1) H$. At this point the mode functions which source the fluctuations of the field can no longer be accurately described by the simple form of noise correlator in the definition of \Eq{eq:langevin}, and hence the stochastic formalism cannot be exactly trusted when $M_i>H$. It is known~\cite{Bunch:1978yq, Birrell:1982ix, Markkanen:2016aes}, however, that the suppression may be further enhanced when $M_i > H$ --- assuming that $M_i$ is constant --- and so we anticipate that further (perhaps fully QFT-theoretic) computations to include a field-dependent effective mass in future work may support our current conjecture beyond this point.

In extended work it would be interesting to study the effect of a number of alternative couplings on the formation of post-inflationary spectator condensates, such as e.g. trilinear couplings (as in the case of the Higgs~\cite{Ema:2017ckf}) and couplings of multiple fields to the inflationary sector (as in `M-flation'~\cite{Ashoorioon:2009wa} or the newly proposed model of `Horizon Feedback Inflation'~\cite{Fairbairn:2017krt}). The abundance of post-inflationary condensates can also be known to affect various models of dark energy~\cite{Glavan:2017jye}, dark matter~\cite{Enqvist:2017kzh}, and the reheating decay efficiency of the inflaton through Higgs thermal blocking~\cite{Freese:2017ace}. Even from these few examples, it is clear that precise numerical predictions provide an important component in providing the initial conditions to model building in the early Universe.

\section*{Acknowledgements}

The author is supported by UK Science and Technology Facilities
Council grant ST/N5044245. Some numerical computations were done on the Sciama High Performance Compute (HPC) cluster which is supported by
the ICG, SEPNet and the University of Portsmouth. The author would also like to thank both Matthew Hull for fruitful discussions and Chris Pattison, Vincent Vennin and David Wands for careful reading and helpful comments on the manuscript.

\newpage

\appendix

\section{Numerical implementation}~\label{sec:Num-implement}

Few analytic solutions to either \Eq{eq:langevin} or \Eq{eq:dist} for $n_{\rm f} \geq 2$ are known to exist, except in the stationary limit of various cases, as given in \Eq{eq:exp-stat-dist}. A robust method for numerical evaluation of a coupled system of Langevin equations of the form in \Eq{eq:langevin} is the modified Improved Euler scheme, introduced in \Ref{2012arXiv1210.0933R}, where it is also proven to exhibit strong first-order convergence. Due to the more complicated potentials studied, a relatively simple implementation of this scheme was developed for the numerical solutions obtained in this work.

The code is written in the python language and achieves runtimes of $\sim$ 5-10 minutes on a standard netbook laptop for $10^4$ realisations of with any potential up to $n_{\rm f} \sim 10$ for $10^5$ \efold{s}. For increased performance, e.g. $n_{\rm f} \sim {\cal O}(100)$ or more, then it is advised to use a computer cluster. The code has also been made publicly available at the following repository: \href{https://github.com/umbralcalc/nfield}{https://github.com/umbralcalc/nfield}. The repository also contains an example script with 5 fields to help the user get started.

In \Fig{fig:pdf-cur-plot1} and \Fig{fig:pdf-cur-plot2} we have plotted some binned realisations of \Eq{eq:langevin} that are used in the code. These plots can also serve as a useful tool to test for numerical convergence, e.g. to check that no arbitrary asymmetry has appeared or if the divergence of the probability current has not vanished due to elevated numerical noise. In such instances, the code may simply be rerun with more realisations to ensure convergence. Even though $10^7$ realisations were used for these plots, numerical noise (and noise from a finite number of samples) still appears for those values of $\boldsymbol{\nabla} \cdot \boldsymbol{J}$ and $\boldsymbol{\nabla} \times \boldsymbol{J}$ which are meant to vanish. Up to this noise amplitude, however, a strong signal can still be seen in $\boldsymbol{\nabla} \times \boldsymbol{J}$ for the $g=10^2g_{\rm dec}$ potential in \Fig{fig:pdf-cur-plot2}, and we leave further improvements to these visualisations for future work.

\clearpage
\bibliographystyle{JHEP}
\bibliography{multispec}

\end{document}

%% file: newcommands.tex
\newcommand{\dd}{\mathrm{d}}
\newcommand{\efold}{$e$-fold}

\newcommand{\beq}{\begin{equation}}
\newcommand{\eeq}{\end{equation}}
\newcommand{\bea}{\begin{eqnarray}}
\newcommand{\eea}{\end{eqnarray}}

\newlength{\wsingfig}
\newlength{\wdblefig}
\newlength{\wquadfig}
\newlength{\wtriplefig}
\newcommand{\Eq}[1]{Eq.~(\ref{#1})}

\newcommand{\Fig}[1]{Fig.~{\ref{#1}}}

\newcommand{\Ref}[1]{Ref.~{\cite{#1}}}

\newcommand{\Sec}[1]{Sec.~\ref{#1}}